\documentclass[a4paper,fleqn,usenatbib]{mnras}

\usepackage{mathptmx}
\usepackage[T1]{fontenc}
\usepackage{ae,aecompl}

\usepackage{graphicx}	% Including figure files

\usepackage{amsmath}	% Advanced maths commands
\usepackage{amssymb}	% Extra maths symbols

\newcommand{\degree}{$^{\rm o}$}	% per cm-squared

\title[SRT observations of the galaxy cluster 3C\,129]{Sardinia Radio Telescope wide-band spectral-polarimetric observations of the galaxy cluster 3C\,129}

\author[M. Murgia et al.]{M. Murgia$^{1}$\thanks{E-mail: matteo@oa-cagliari.inaf.it},
F. Govoni$^{1}$,
E. Carretti$^{1}$, 
A. Melis$^{1}$,
R. Concu$^{1}$, 
A. Trois$^{1}$, 
F. Loi$^{1, 2}$, 
\newauthor 
V. Vacca$^{1}$,  
A. Tarchi$^{1}$, 
P. Castangia$^{1}$,
A. Possenti$^{1}$, 
A. Bocchinu$^{1}$, 
M. Burgay$^{1}$, 
\newauthor 
S. Casu$^{1}$, 
A. Pellizzoni$^{1}$, 
T. Pisanu$^{1}$, 
A. Poddighe$^{1}$, 
S. Poppi$^{1}$,
N. D'Amico$^{1,2}$,
\newauthor
M. Bachetti$^{1}$, 
A. Corongiu$^{1}$, 
E. Egron$^{1}$, 
N. Iacolina$^{1}$, 
A. Ladu$^{1}$, 
P. Marongiu$^{1}$, 
 \newauthor
C. Migoni$^{1}$, 
D. Perrodin$^{1}$, 
M. Pilia$^{1}$, 
G. Valente$^{1}$, 
and  G. Vargiu$^{1}$
\\
$^{1}$INAF\,-\,Osservatorio Astronomico di Cagliari, Via della Scienza 5 - I-09047 Selargius (CA), Italy\\
$^{2}$Dipartimento di Fisica, Universit\`a di Cagliari, Strada Prov.le Monserrato-Sestu Km 0.700, I-09042 Monserrato (CA), Italy.\\
}

\date{Accepted XXX. Received YYY; in original form ZZZ}

\pubyear{2016}

\begin{document}
\label{firstpage}
\pagerange{\pageref{firstpage}--\pageref{lastpage}}
\maketitle

% Abstract of the paper
\begin{abstract}
We present new observations of the galaxy cluster 3C\,129 obtained with the Sardinia Radio Telescope in the frequency range 6000$-$7200\,MHz, with the aim to image the large-angular-scale emission at high-frequency of the radio sources located in this cluster of galaxies. The data were acquired using the recently-commissioned ROACH2-based backend to produce full-Stokes image cubes of an area of 1\degree$\times$1\degree~centered on the radio source 3C\,129. We modeled and deconvolved the telescope beam pattern from the data. We also measured the instrumental polarization beam patterns to correct the polarization images for off-axis instrumental polarization. Total intensity images at an angular resolution of 2.9\arcmin~were obtained for the tailed radio galaxy 3C\,129 and for 13 more sources in the field, including 3C\,129.1 at the galaxy cluster center. These data were used, in combination with literature data at lower frequencies, to derive the variation of the synchrotron spectrum of 3C\,129 along the tail of the radio source. If the magnetic field is at the equipartition value, we showed that the lifetimes of radiating electrons result in a radiative age for 3C\,129 of $t_{\rm syn}\simeq 267\pm 26$ Myrs. Assuming a linear projected length of 488\,kpc for the tail, we deduced that 3C\,129 is moving  supersonically with a Mach number of $M=v_{\rm gal}/c_{\rm s}=1.47$.  Linearly polarized emission was clearly detected for both 3C\,129 and 3C\,129.1. The linear polarization measured for 3C\,129 reaches levels as high as 70\% in the faintest region of the source where the magnetic field is aligned with the direction of the tail.
\end{abstract}

% Select between one and six entries from the list of approved keywords.
% Don't make up new ones.

%%%%%%%%%%%%%%%%%%%%%%%%%%%%%%%%%%%%%%%%%%%%%%%%%%

%%%%%%%%%%%%%%%%% BODY OF PAPER %%%%%%%%%%%%%%%%%%

\section{Introduction}

The Sardinia Radio Telescope (SRT; Grueff et al. 2004) is a new 64-m diameter 
radio telescope located at San Basilio, about 35\,km north of Cagliari, Italy. The telescope is designed for astronomy, geodesy, and space
science, either as a single dish or as part of European and global networks.
The SRT has a shaped Gregorian optical 
configuration with a 7.9-m secondary
mirror and supplementary Beam-WaveGuide (BWG) mirrors. 
With six focal positions
(primary, Gregorian, and four BWGs), the SRT is capable of allocating up to 20 
receivers. Eventually, once all of the planned devices are installed, it will operate
with high efficiency in the 0.3-116 GHz frequency range.  
One of the most advanced technical 
features of the SRT is its active surface:
the primary mirror is composed of 1008 panels supported by 
electromechanical actuators that are
digitally controlled to compensate for gravitational 
deformations.

The antenna officially opened on September 30th 2013, upon completion of the technical
commissioning phase (Bolli et al. 2015). In its first light configuration, the SRT is equipped with
three receivers: a 7-beam K-band (18-26.5 GHz) receiver (Gregorian focus; Orfei et al. 2010), 
a single-feed C-band (5.7-7.7 GHz) receiver (BWG focus), 
and a coaxial dual-feed L/P-band (0.305-0.41/1.3-1.8 GHz) receiver (primary focus; Valente et al. 2010). 
The technical commissioning phase was followed by the Astronomical Validation activity aimed at validating the 
telescope for standard observing modes and at transforming the SRT from a technological project into a real
general-purpose astronomical observatory. The Astronomical Validation results are presented by Prandoni et al. (in preparation).

The suite of backends currently available on site includes the
recently-commissioned ROACH2 FPGA board-based\footnote{The Reconfigurable Open Architecture Computing Hardware (ROACH) processing board is developed by
the Center for Astronomy Signal Processing and Engineering Research, see http://casper.berkeley.edu.The ROACH2 is a stand-alone FPGA-based board that represents the successor to the original ROACH board.}  backend SARDARA (SArdinia Roach2-based Digital Architecture for Radio Astronomy; Melis et al., in preparation). 

In this work, we exploited for the first time the capabilities of the
SARDARA backend in spectral-polarimetric mode to perform total intensity and 
polarization observations in C-band. For this purpose, we observed  the galaxy cluster 3C\,129 field. 3C\,129 is a luminous X-ray galaxy 
cluster at redshift of $z=0.0208$ (Spinrad 1975), located close to the Galactic plane. 
We selected this field because of the homonymous radio galaxy 3C\,129, 
whose spectral and polarization properties have been very well studied in the literature.
Indeed, these observations serve as an important term of reference to assess the reliability of the calibration of the new 
spectral-polarimetric data acquired with the SRT. Moreover, they provide new information on the high-frequency 
large angular scale emission of the radio sources located in this cluster of galaxies.

Like many other radio sources hosted in galaxy clusters, 3C\,129 has a distorted head-tail morphology
where the jets are bent into a tail due to the ram-pressure caused by the motion of the parent galaxy in 
the intracluster medium. The radio galaxy 3C\,129 was among the first radio galaxies discovered to have a 
head-tail structure. It is located at the cluster periphery and  
was identified with an E-galaxy by Hill \& Longair (1971). Since then, it has been studied
extensively in radio by several authors (e.g. Miley 1973, Perley
\& Erickson 1979, Owen et al. 1979, van Breugel 1982,  
Downes 1984, J{\"a}gers 1987, Feretti et al. 1998, Taylor et al 2001).
The source stretches over $\sim$20\arcmin~in length, and
the spectacular structure of its narrow-angle tail 
has been shown at low frequencies with Very Large Array (VLA) and Giant Metrewave Radio Telescope (GMRT) observations
(e.g. Kassim et al. 1993, Lane et al. 2002, Lal \& Rao 2004).
The importance of mapping the radio galaxy 3C\,129 with a single dish is that interferometers go into the 
technical problem of the missing zero spacing. Indeed, they
filter out structures larger than the angular 
size corresponding to their shortest spacing, limiting
the synthesis imaging of extended structures. 
Single dish telescopes are optimal for recovering 
all of the emission on large angular scales, especially at high frequencies. Although single dishes
typically have a low resolution, the 3C\,129 tail stretches
enough to be well resolved at the resolution of about 2.9\arcmin~of the SRT
in C-band. 

In addition to the head-tail 3C\,129, we mapped other radio sources in the field, 
including the radio source 3C\,129.1 near the projected 
center of the X-ray galaxy cluster. This radio source also has dual radio jets
but it is weaker and smaller than 3C\,129, and the radio jets extend 
over about  2\arcmin~(e.g. Downes 1984, J{\"a}gers 1987, 
Kassim et al. 1993). 

The paper is organized as follows, in Sect.\,2, we illustrate the SRT observations. 
In Sect.\,3, we present the data reduction.
In Sect.\,4, we show the total intensity and polarization imaging.
In Sect.\,5, we present the spectral analysis. 
In Sect.\,6, we present the polarization analysis. 
Finally, in Sect.\,7 we draw our summary. In Appendix, we provide details about the data reduction. 

Throughout this paper, we assume a $\Lambda$CDM cosmology with
$H_0$ = 71 km s$^{-1}$Mpc$^{-1}$, $\Omega_m$ = 0.3, and $\Omega_{\Lambda}$ = 0.7.
At the distance of 3C\,129, 1\arcsec~corresponds to 0.415 kpc.

\section{Observations}

We observed a $1^\circ\times1^\circ$ area centered at the radio
source 3C~129 (J2000 coordinates: RA 04$^{\rm h}$49$^{\rm m}$03$^{\rm s}$ and DEC +45\degree01\arcmin43\arcsec),
using the C--band receiver of the SRT. The receiver is a circular polarization system, 
installed at the BWG focus of the telescope, covering the band 5700-7700\,MHz, with system temperature $T_{\rm sys} \simeq 30$\,K.
The SRT active surface was used to correct the primary mirror shape and keep the gain curve nearly flat 
at all elevations. The data were acquired with the digital correlator SARDARA. The autocorrelations of the two polarizations 
($RR^*$ and $LL^*$) and their complex cross-product ($RL^*$) were recorded for full Stokes information.
We used the correlator configuration with 1500\,MHz bandwidth and 1024 frequency channels. Each channel was of 1.46\,MHz in width.
We observed a bandwidth of 1250\,MHz at a central frequency of 6600\,MHz. The angular resolution is 2.9\arcmin. 

The area was mapped with two sets of orthogonal scans along RA and DEC,
with scan speed of $6^\circ$/min and 0.7\arcmin~scan spacing to ensure Nyquist sampling 
of the beam. The data sampling rate was 10\,ms for an angular sampling rate along the scan direction of 3.6\arcsec.
Band-Pass, flux density, and polarization angle calibration were performed with the source 3C\,286. We used 3C\,84 for instrumental
polarization leakage calibration. Mapping in AZ-EL (instrument) reference frame of the sources 
3C\,273, 3C\,345, 3C\,454.3, and 3C\,84 was done to measure the beam pattern 
for all Stokes parameters to correct both the beam shape in Stokes I and clean
the off-axis instrumental polarization in Stokes Q and U.
The sources 3C\,138 and NGC\,7027 were observed as secondary calibrators to cross-check the accuracy 
of the polarization position angle calibration and the correction of the instrumental polarization, 
respectively.

A summary of the observations is listed in Tab. 1.

\begin{table}
\label{Observations}
\caption{SRT observations details.}             
\centering          
\begin{tabular}{c c c c }     
\hline\hline
Source          & Date            & OTF scan axis                        & Elevation \\	  
\hline\hline
3C\,129         &26 Jul 2015      & 4$\times$ (RA-DEC)                   & 18\degree~$-$55\degree\\
3C\,84          &26 Jul 2015      & 1$\times$ (RA-DEC)                   & 42\degree~$-$44\degree\\
3C\,84          &26 Jul 2015      & 2$\times$ CROSS                      & 44\degree\\
3C\,138         &26 Jul 2015      & 2$\times$ CROSS                      & 56\degree\\
3C\,286         &26 Jul 2015      & 3$\times$ CROSS                      & 72\degree\\

\hline

3C\,129         &02 Aug 2015      & 3$\times$ (RA-DEC)+1DEC              & 21\degree~$-$53\degree\\
3C\,84          &02 Aug 2015      & 1$\times$ (RA-DEC)                   & 40\degree~$-$41\degree\\
3C\,84          &02 Aug 2015      & 2$\times$ CROSS                      & 43\degree\\
3C\,138         &02 Aug 2015      & 2$\times$ CROSS                      & 54\degree\\
3C\,286         &02 Aug 2015      & 3$\times$ CROSS                      & 67\degree\\
3C\,273         &02 Aug 2015      & 2$\times$ (AZ-EL)                    & 52\degree\\

\hline

3C\,129         &10 Sep 2015      & 3$\times$ RA +2$\times$ DEC          & 39\degree~$-$66\degree\\
3C\,84          &10 Sep 2015      & 3$\times$ CROSS                      & 53\degree\\
3C\,138         &10 Sep 2015      & 10$\times$ CROSS                     & 66\degree\\
3C\,286         &10 Sep 2015      & 7$\times$ CROSS                      & 43\degree\\
\hline
3C\,454.3       &31 Jul 2015      & 1$\times$ (AZ-EL)                    & 40\degree~$-$42\degree\\
\hline
3C\,454.3       &25 Oct 2015      & 4$\times$ (AZ-EL)                    & 23\degree~$-$42\degree\\

\hline
3C\,454.3       &07 Aug 2015      & 3$\times$ (AZ-EL)                    & 27\degree~$-$40\degree\\

\hline
3C\,345         &15 Nov 2015      & 2$\times$ (AZ-EL)                    & 64\degree~$-$73\degree\\
\hline

3C\,129         &12 Dec 2015      & 1$\times$ (RA-DEC)                  & 39\degree~$-$66\degree\\
3C\,84          &12 Dec 2015      & 1$\times$ (RA-DEC)                  & 67\degree~$-$73\degree\\
3C\,138         &12 Dec 2015      & 2$\times$ CROSS                     & 67\degree\\
3C\,286         &12 Dec 2015      & 4$\times$ CROSS                     & 52\degree\\

\hline

NGC\,7027       &06 Feb 2016      & 2$\times$ (RA)                      & 57\degree~$-$67\degree\\
3C\,84          &06 Feb 2016      & 4$\times$ CROSS                     & 40\degree\\
3C\,138         &06 Feb 2016      & 3$\times$ CROSS                     & 49\degree\\
3C\,286         &06 Feb 2016      & 7$\times$ CROSS                     & 25\degree~$-$30\degree\\

\hline\hline
\end{tabular}   
\end{table}

\section{Data reduction}

\begin{figure}
\begin{center}
\includegraphics[width=9cm]{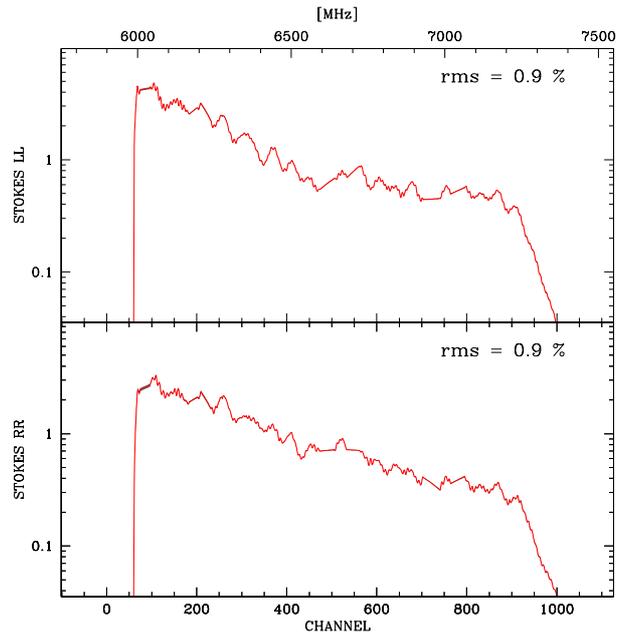}
\end{center}
\caption{Normalized band-pass for the left (top panel) and right (bottom panel) circular polarizations. Solutions were interpolated over flagged channels.}
\label{bpass}
\end{figure}

Data reduction was performed with the proprietary Single-dish Spectral-polarimetry 
Software (SCUBE; Murgia \& Govoni). In the following, we present the steps of the calibration procedure.

\begin{figure}
\begin{center}
\includegraphics[width=9cm]{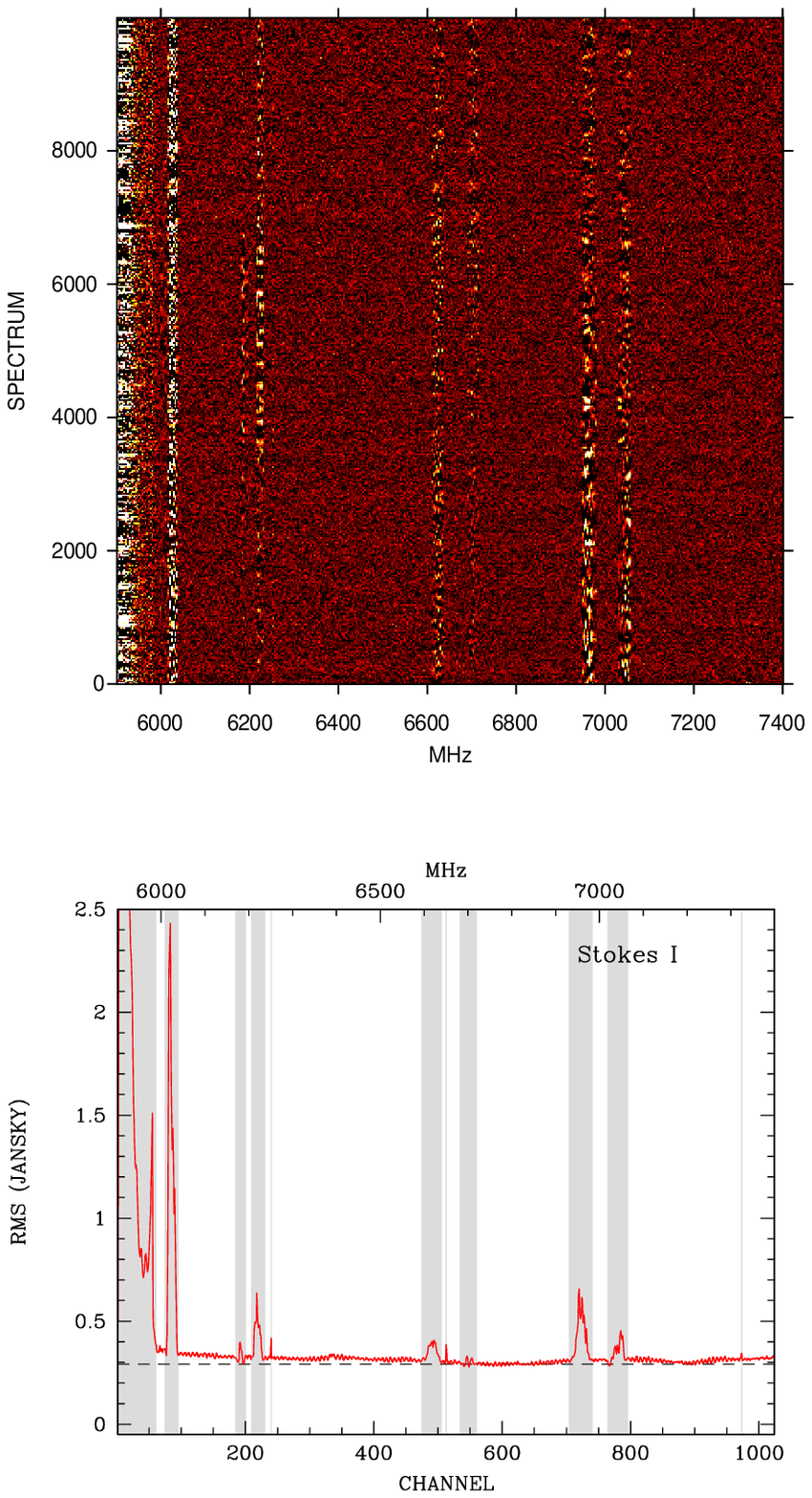}
\end{center}
\caption{Spectrograms showing the typical RFI scenario for Stokes I. Top panel: spectrogram composed by about 10,000 spectra measured over a period of about 2 hours during 02\,Aug\,2015 session in a region of the 3C\,129 field relatively free of strong sources. The baseline was subtracted from each scan channel-by-channel by mean of a linear fit. We corrected for the band-pass and data were converted to Jansky. The color map ranges from -1 to 1 Jy. Bottom panel: channel noise rms. The horizontal dashed line represents the expected reference thermal noise level of 0.29 Jy traced assuming $T_{\rm sys}$=30 K and an antenna gain of $\Gamma$=0.6 Jy/K. The vertical shaded bands identify the blocks of channels we flagged from the data.}
\label{spectrogram}
\end{figure}

\subsection{Flux density calibration and RFI flagging}
Band-Pass and flux density calibration were obtained using 3C\,286 observations
assuming the flux density scale of Perley \& Butler (2013).
RFI flagging was performed iteratively to
minimize the impact of RFI on flux density calibration. 
After a first band-pass and flux density calibration cycle, RFI were flagged using observations of a 
cold part of the sky without obvious sources.
The flagged data were then used to repeat the band-pass and the flux density calibration
for a finer RFI flagging.
The procedure was iterated a few times, eliminating all of the most obvious RFI.

\begin{figure}
\begin{center}
\includegraphics[width=9cm]{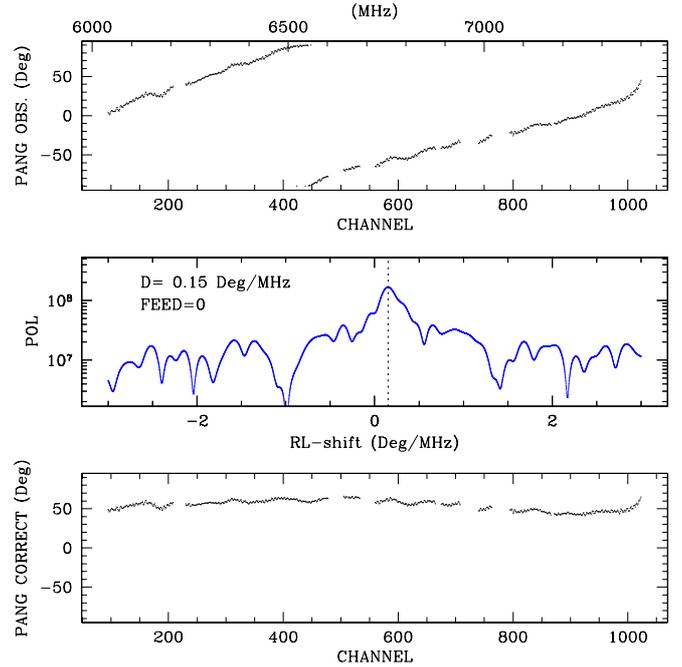}
\end{center}
\caption{Example of the calibration of the phase shift between the R and L polarizations. Top: observed polarization angle. Middle: polarization synthesis. The peak polarization is found at a RL-shift of 0.15\,deg/MHz. Bottom: corrected 
polarization angle. }
\label{RLshift}
\end{figure}

\begin{figure}
\begin{center}
\includegraphics[width=8cm]{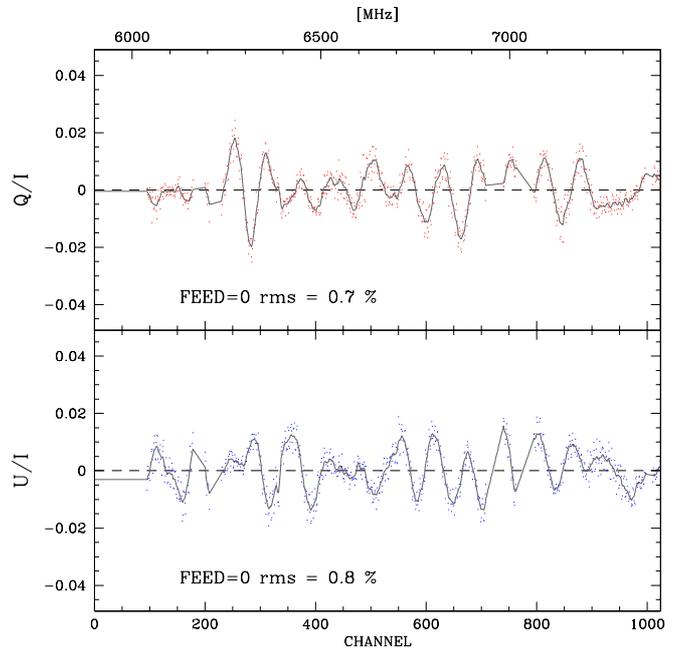}
\end{center}
\caption{On-axis polarization leakage terms determined from 3C\,84 scans for 02\,Aug\,2015 session here are represented as the ratio of Q/I and U/I on top and bottom panel, respectively. The continuous lines is a data smoothing over a scale of 4 channels.}
\label{onaxispol}
\end{figure}

The normalized band-pass for the left and right circular polarizations is shown in Fig.\,1. 
The solution obtained from all of the calibration scans on 3C\,286 were averaged and then 
interpolated on RFI flagged frequency channels and their rms dispersion (i.e. the scatter of the solutions from all calibration scans) is below 1\%.

\begin{figure*}
\begin{center}
\includegraphics[width=18cm]{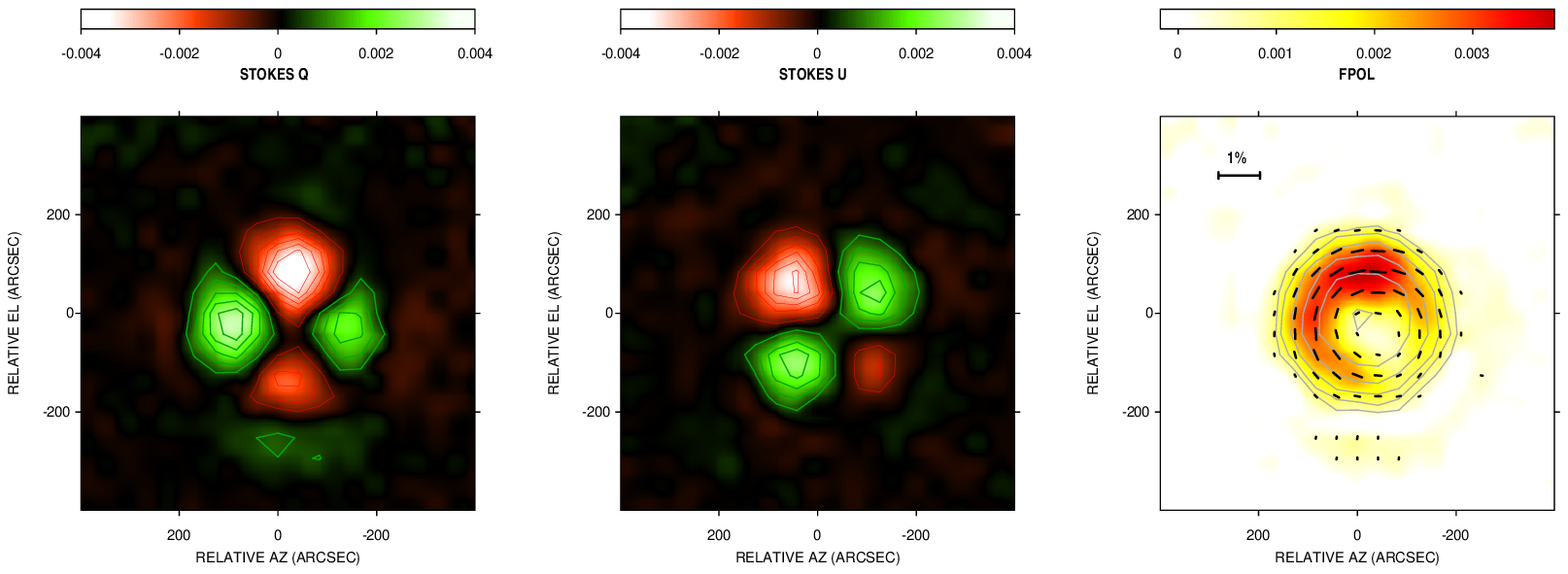}
\end{center}
\caption{Off-axis polarization between 6000 and 7200\,MHz derived from 3C\,84 imaging. Units are relative to the Stokes I peak. The left and middle panels show the Stokes Q and U patterns corrected for the on-axis instrumental polarization. The right panel shows the off-axis fractional polarization vectors overlaid on the image of polarized intensity divided by the total intensity peak. Contours refer to Stokes I. The off-axis fractional polarization level is about 0.3$-$0.4\%. Polarization angles are measured counter-clockwise starting from the vertical axis.}
\label{offaxispol}
\end{figure*}

We applied the gain-elevation curve correction to account for the gain variation  
with elevation due to the telescope structure gravitational stress change. 
We assumed the values measured during the Astronomical Validation (Prandoni et al. 2016, in prep.)

\begin{figure}
\begin{center}
\includegraphics[width=9cm]{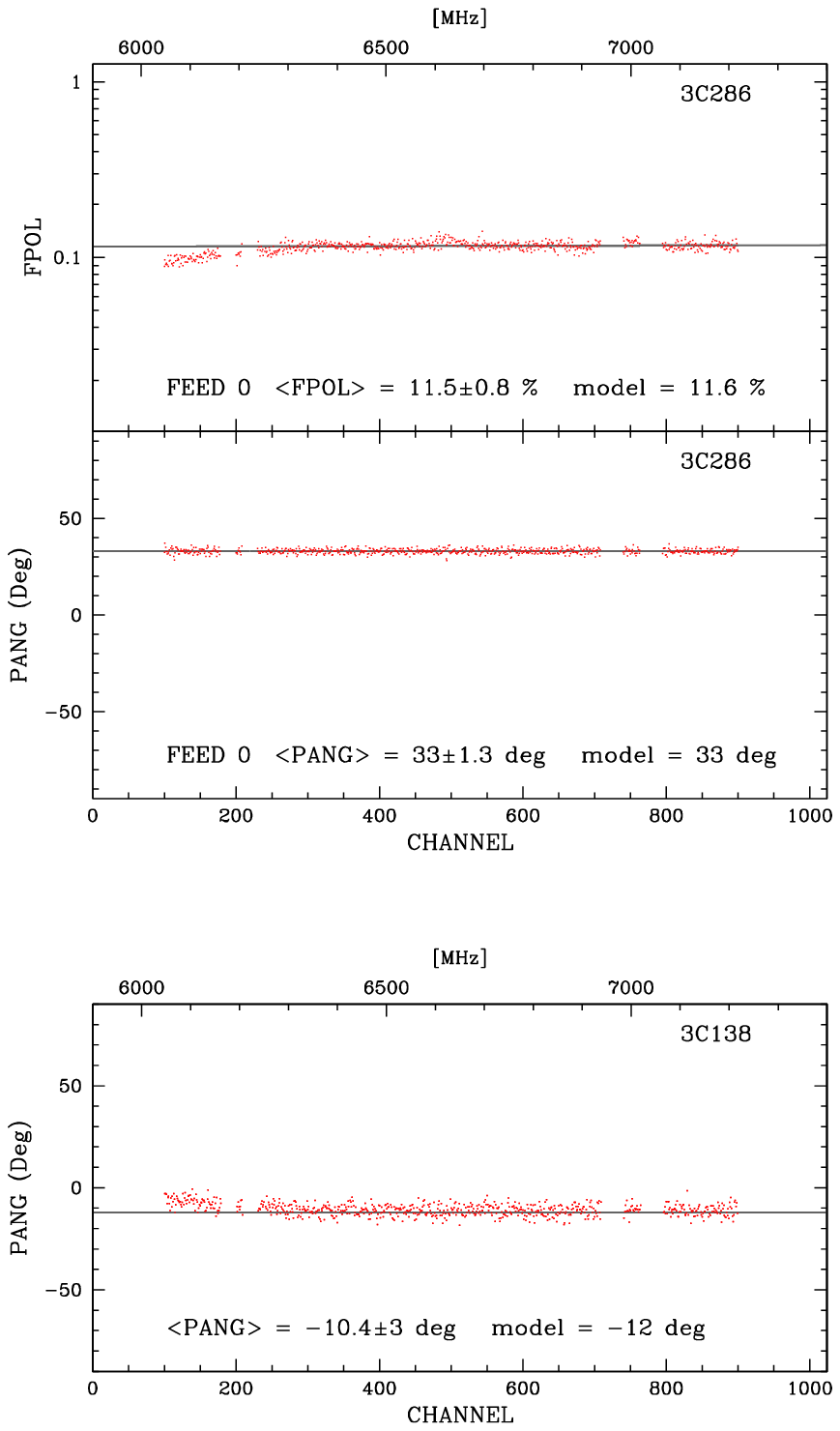}
\end{center}
\caption{Top: calibrated fractional polarization corrected for the on-axis instrumental polarization (upper panel) and polarization position angle (lower panel) for 3C\,286 observed on 02\,Aug\,2015 are shown as red dots. The continuous lines represent the expected values based on the model of Perley \& Butler (2013). The comparison between data and model statistics is calculated for the inner part of the bandwidth between channels 100 and 900. Bottom: cross-check of the polarization position angle calibration on the secondary calibrator 3C\,138 observed in the same session.}
\label{pola}
\end{figure}

Fig.\,2, top panel, shows the typical RFI scenario of our observations. 
We selected about 10,000 spectra taken over 2 hour observations on 2 August 2015 
in a region of the 3C\,129 field with no strong sources. Spectra were corrected for the band-pass and a linear baseline was subtracted 
from each scan for each frequency channel. 

The rms noise measured for each channel is plotted in Fig.\,\ref{spectrogram}, bottom panel. 
Where free from RFI, the noise is in good agreement with the theoretical value  
of 0.29\,Jy expected for an integration time of 10\,ms, a bandwidth of 1.46\,MHz per channel, 
a $T_{\rm sys}$=30 K, and a gain of $\Gamma$=0.6 Jy/K.

\subsection{Polarization calibration}
The polarization calibration was applied in three stages:
correction for the phase difference between R and L; correction for the leakage of Stokes I into Q and U; 
calibration of the absolute position of the polarization angle.

Phase difference between R and L can be added at any stage of the system (e.g. any device in the signal path, or 
any difference of the cable length of the two polarization paths) that results in a rotation of the measured polarization angle
for a system using circular polarizations. In addition, we found that the time calibration of 
the two samplers of the backend (one each polarization) could be affected by an offset of an integer multiple of the
sampling time. This results in a time delay of the two polarizations, and, in turn, a phase difference changing 
with a linear behavior through the frequency band. To account for that, observations of 
polarized calibration sources were used and the phase corrected to match that of the source.
Typically we measure phase difference variations of about 0.1\,deg/MHz (see Fig.\,\ref{RLshift}).
The phase difference can change any time the correlator is configured, but we checked it is stable 
in--between reconfigurations. We thus observed a polarization calibrator at the beginning of each observing session or any time
a reconfiguration of the backend was required.

The on-axis instrumental polarization 
was measured with observations of the unpolarized 
source 3C\,84 (flux density $S_{\rm 6600\,MHz}\simeq 27$ Jy at the time of our observations), 
and is reported in Fig.\,4 where the fractional instrumental polarization of the source is shown. 
The leakage is in 
general better than 2\% across the band, with a rms scatter of 0.7 - 0.8 \% with variations on scales of some 100\,MHz. 
Because of the slow variation with frequency we smoothed on a scale of 4 channels ($\Delta\nu\simeq 6$\,MHz) to reduce their error.

We also corrected for the off-axis instrumental polarization. Fig.\,5 shows the instrumental polarization beam for Stokes Q and U we have measured
with observations of 3C\,84, after correction for the on-axis instrumental polarization. 
The plots show the typical butterfly pattern (e.g. see Gregorini et al. 1992, Carretti et al. 2004, O'Dea et al. 2007).
The off-axis instrumental polarization level compared to the Stokes I peak is 0.3\%. 
This contribution can strongly affect the quality of polarization data if bright sources are present in the image.
We corrected for them by deconvolving these instrumental beam patterns in 
the instrument reference alt-azimuth frame (see Sect. 4.2).

As a final step, we calibrated the absolute position of the polarization angle for each frequency channel 
using the primary polarization calibrator 3C\,286, assuming a polarization angle of 33\degree~(Perley \& Butler 2013) through the entire frequency band.
The calibrated 3C\,286 fractional polarization and position angle, corrected for the 
leakage and the parallactic position angle, are shown in Fig.\,6. The fractional polarization of 3C\,286 is in excellent agreement with the value of 11.6\% reported in 
Perley \& Butler (2013), and the scatter of the position angle solution is about 1\degree.

We checked the accuracy of the polarization angle calibration by applying the entire procedure to
3C\,138, which we observed as an independent secondary polarization calibrator during each session. We found that the calibrated polarization angle for 3C\,138 is consistent
 within the errors with the expected value of -12\degree~reported by Perley \& Butler (2013), as shown in the bottom panel of Fig.\,6.
 
The accuracy of our determinations of the fractional polarization was checked by applying the calibration procedure to the radio source NGC\,7027, which we 
assumed to be unpolarized. Before the correction for the leakage terms, we found that the on-axis fractional polarization was of $2\pm 0.1$\%. 
After correcting for the on-axis leakage terms calibrated with 3C\,84, the residual instrumental polarization observed for NGC\,7027 was found to be $< 0.1$\% (3$\sigma$ upper limit).

\section{Results}

\subsection{Total intensity imaging}

The imaging was performed in SCUBE by subtracting the baseline from the calibrated telescope scans and by projecting the data into a regular 3-dimensional
 grid with a spatial resolution of 42\arcsec/pixel, which is enough in our case to sample the beam FWHM with four pixels. 

We were primarily interested in imaging the discrete radio sources in the 3C129 field, and the mapped area was designed to have the target sources at the center. Thus, the baseline was subtracted starting from scratch by a linear fit involving only the 10\% of data at the beginning and the end of each scan. By doing this, we filtered out real sky signals (like e.g. the atmospheric emission, the Cosmic Microwave Background radiation, or the Galactic foreground), those fluctuations whose surface brightness gradient is uniform on angular scales smaller than about 1\degree. The baseline removal was performed channel-by-channel for each scan.
All frequency cubes obtained by gridding the scans along the two orthogonal axes (RA and DEC) were then stacked together to produce full-Stokes I, Q, U images 
of an area of 1 square degree centered on the radio source 3C\,129. In the combination, the individual image cubes 
were averaged and de-stripped by mixing their Stationary Wavelet Transform (SWT) coefficients (hereafter SWT stacking) as described in the Appendix.
We then used the higher Signal-to-Noise (S/N) image cubes obtained from the SWT stacking as a prior model to refine the baseline fit. 
The baseline subtraction procedure was repeated including all data and not just the 10\% from the beginning and the end of each scan. In this way, we were able to 
improve the baseline subtraction for those scans intercepting a radio source (or a source of RFI) located just at the edge of the map. 
In the refinement stage, the baseline was represented with a 2nd order polynomial whose coefficients were found by minimizing the difference between the model and the data using a least-squares fit (see Fig.\,\ref{baselineref}).
A new SWT stacking was then performed and the process was iterated a few more times until the convergence was reached.

\begin{figure*}
\begin{center}
\includegraphics[width=16cm]{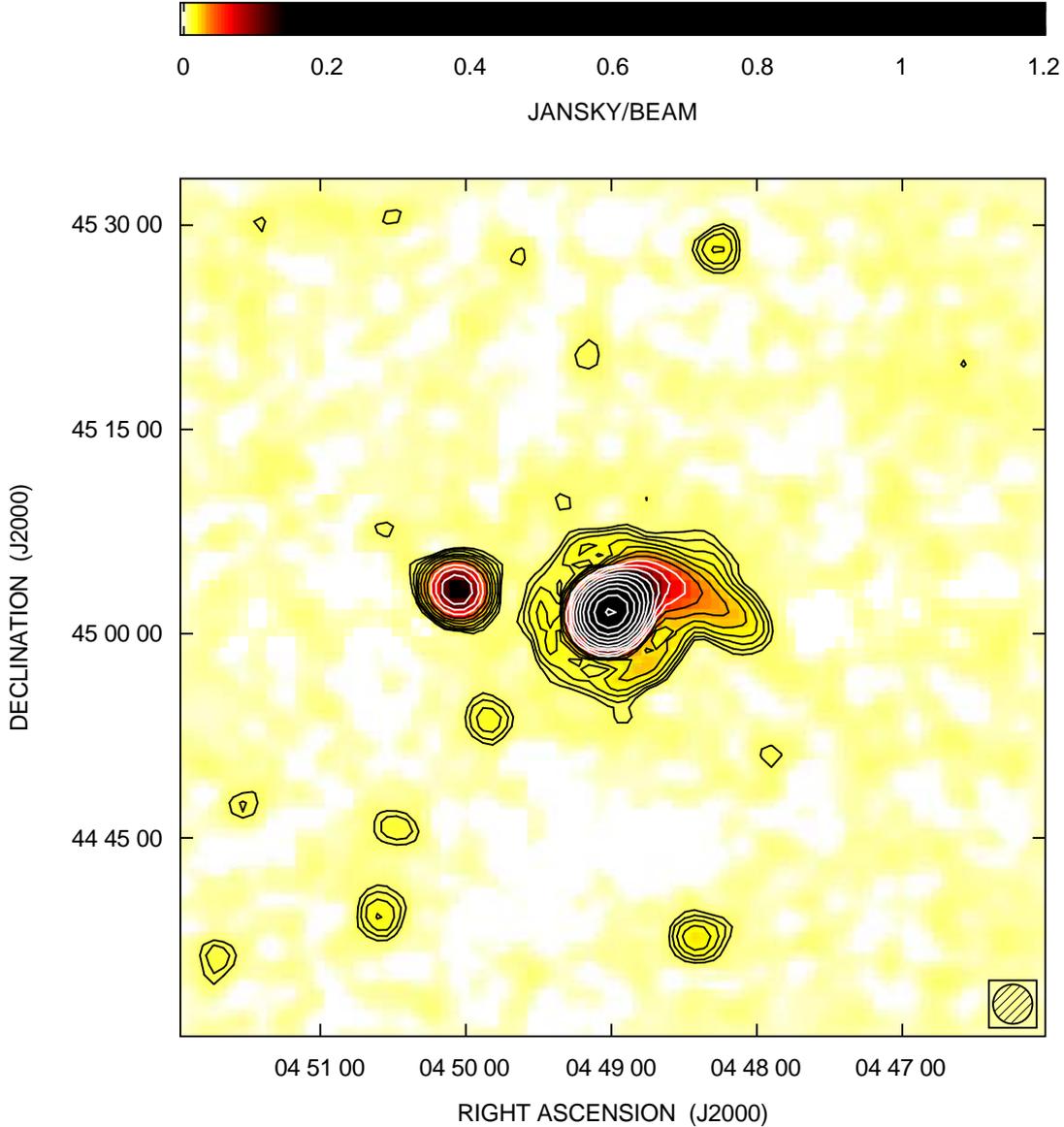}
\end{center}
\caption{SRT total intensity ``dirty map'' of the galaxy cluster 3C\,129 resulting from the spectral average of the bandwidth between 6000 and 7200\,MHz. The noise level is 1.5 mJy/beam. The FWHM beam is 2.9\arcmin, as indicated in the bottom-right corner. Contour levels start at 3$\sigma$ and increase by a factor of $\sqrt{2}$.}
\label{3C129_dirty}
\end{figure*}

Fig.\ref{3C129_dirty} shows the total intensity image obtained averaging all frequency channels from 6000\,MHz to to 7200\,MHz.
We reached a final noise level of 1.5 mJy/beam, 2.9\arcmin~FWHM beam. In Fig.\ref{3C129_dirty} we traced the total intensity contour levels that start from 3$\sigma$ and increase by a factor of $\sqrt{2}$. 

The narrow head-tail radio galaxy 3C\,129 dominates the central part of the image with a peak brightness of about 1.2 Jy/beam. The source's tail extends towards 
west for more than 10\arcmin~before disappearing below the image noise floor. 
With a peak brightness of 142 mJy/beam, 3C\,129.1 is the second brightest radio source in the field. It is located in projection close to 
the cluster center; it is also a narrow angle head-tail radio galaxy but is only slightly resolved in the SRT image.

By comparing the SRT image with other images at lower frequencies taken from the literature, we confirmed the detection of additional 12 fainter sources in the field, with flux densities ranging from 4.1 to 17.7 mJy at 6600\,MHz (see Sect.\ref{specanal}).

\subsection{Beam pattern modeling and de-convolution}

%%%% https://ned.ipac.caltech.edu/level5/March01/Andernach/Ander2.html#2.1

\begin{figure*}
\begin{center}
\includegraphics[width=16cm ]{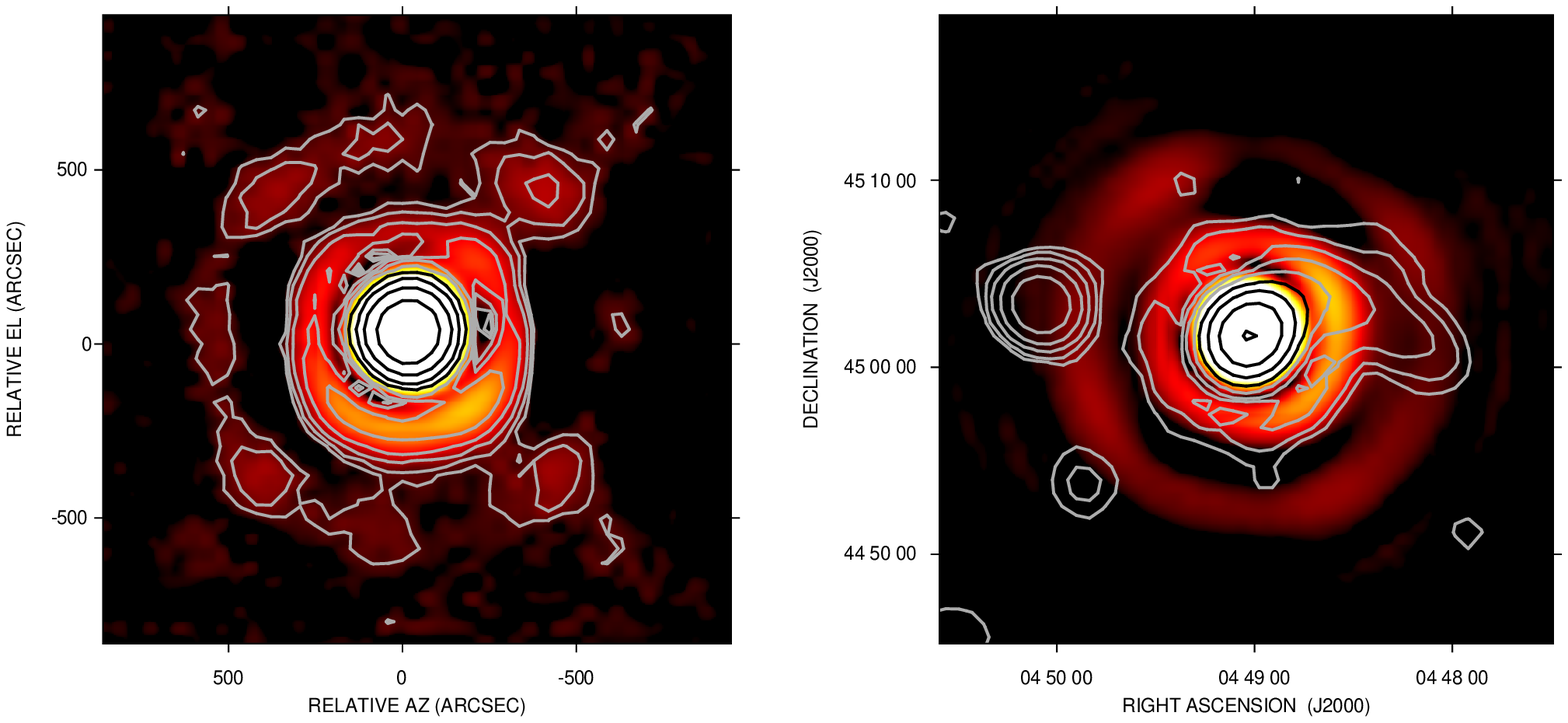}
\end{center}
\caption{Left: SRT beam pattern model at C band in the horizontal AZ-EL frame. Contour levels range from 0 to -30 dB at steps of -3 dB. Right: beam pattern model re-projected into the equatorial RA-DEC frame and averaged over the position angles of the 3C\,129 scans. The contours of the 3C\,129 ``dirty map'' are superimposed. Levels start at 4.5 mJy/beam and increase by a factor of 2. Note that the brightest part of the asymmetric first side-lobe of the beam is oriented towards the tail of the radio galaxy. The beam model shown in the left panel, represents an average over an elevation range very similar to that covered for 3C\,129.}
\label{beam}
\end{figure*}

The dynamic range, defined as the peak-to-noise ratio, expected for the image shown in Fig.\ref{3C129_dirty} is about 800. However, close to
the head of 3C\,129, the dynamic range is clearly limited by the telescope beam pattern rather than by the image sensitivity. In fact, the first side lobe of the SRT
beam, whose nominal level is -20 dB, is obvious around 3C\,129 so that the image should be better interpreted as the ``dirty map'' resulting
from the convolution of the telescope beam pattern with the true sky brightness distribution.

Modeling and de-convolution of the telescope beam pattern are therefore necessary for recovering the expected dynamic range.

We derived a model of the SRT beam pattern by performing a set of OTF scans of the bright quasars 3C\,454.3, 3C\,345, and 3C\,273 (see Tab.\,1). We used the same frontend and backend
setups as for the 3C\,129 field, except that maps were performed in the horizontal frame, rather than in the equatorial frame, by alternating 
scans along the direction of azimuth and elevation. A total of 48 frequency cubes were combined to produce an average beam model. However, we also grouped the
scans into three distinct coarse elevation bins having an equal width of 20\degree, centered at $EL=$30\degree, 50\degree, and 70\degree, to construct an elevation-dependent beam model.

% (20\degree$-$40\degree, 40\degree$-$60\degree, and 60\degree$-$80\degree)
\begin{figure*}
\begin{center}
\includegraphics[width=16cm]{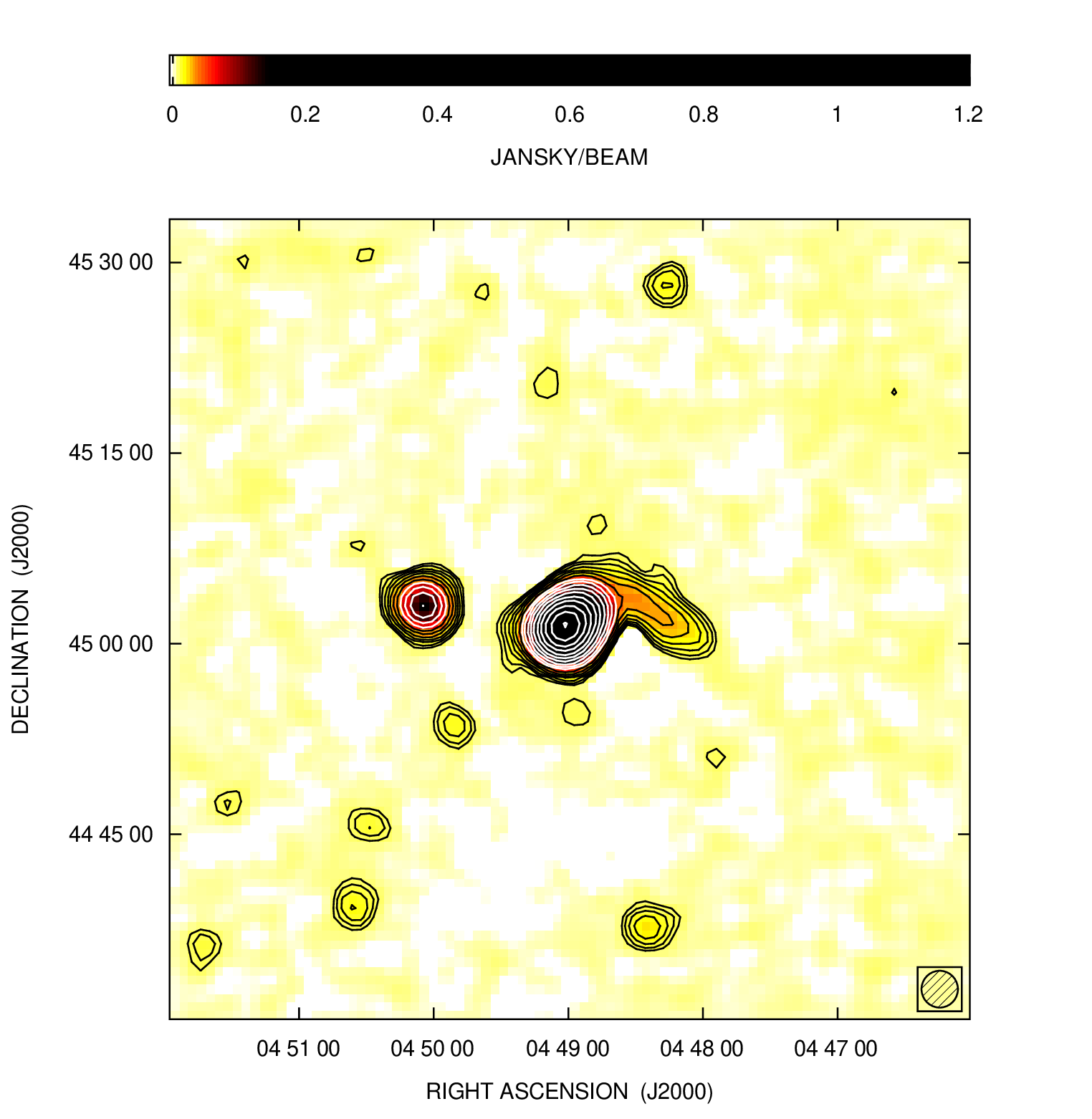}
\end{center}
\caption{SRT total intensity CLEANed image of the galaxy cluster 3C\,129 resulting from the spectral average of the bandwidth between 6000 and 7200\,MHz. The SRT beam pattern was deconvolved from each individual map and the clean components were restored into the residual image with a circular Gaussian with FWHM of 2.9\arcmin. The noise level is 1.5 mJy/beam. The FWHM beam is indicated in the bottom-right corner. Contour levels start at 3$\sigma$ and increase by a factor of $\sqrt{2}$.}
\label{3C129_field_cleaned}
\end{figure*}

In the left panel of Fig.\,\ref{beam}, we show the image of the SRT beam pattern after the SWT stacking of all scans, from all elevation bins, and the averaging of all spectral channels. The main lobe can be considered to be
circularly symmetric Gaussian with FWHM=2.9\arcmin. We have enough S/N to detect both the
first and the second beam side lobes, respectively at -20 and -30 dB. The first side lobe is clearly asymmetric, since it is much brighter below the main lobe. 
The asymmetry decreases as elevation increases, and the beam is nearly symmetric around $EL\simeq 60$\degree.
The most striking features associated with the second side lobe are the four spikes seen at the tips of a cross tilted by 45\degree. These originate from the blockage caused by the quadripode holding sustaining the secondary mirror in the path of light between the source and the primary mirror.

\begin{figure*}
\begin{center}
\includegraphics[width=16cm]{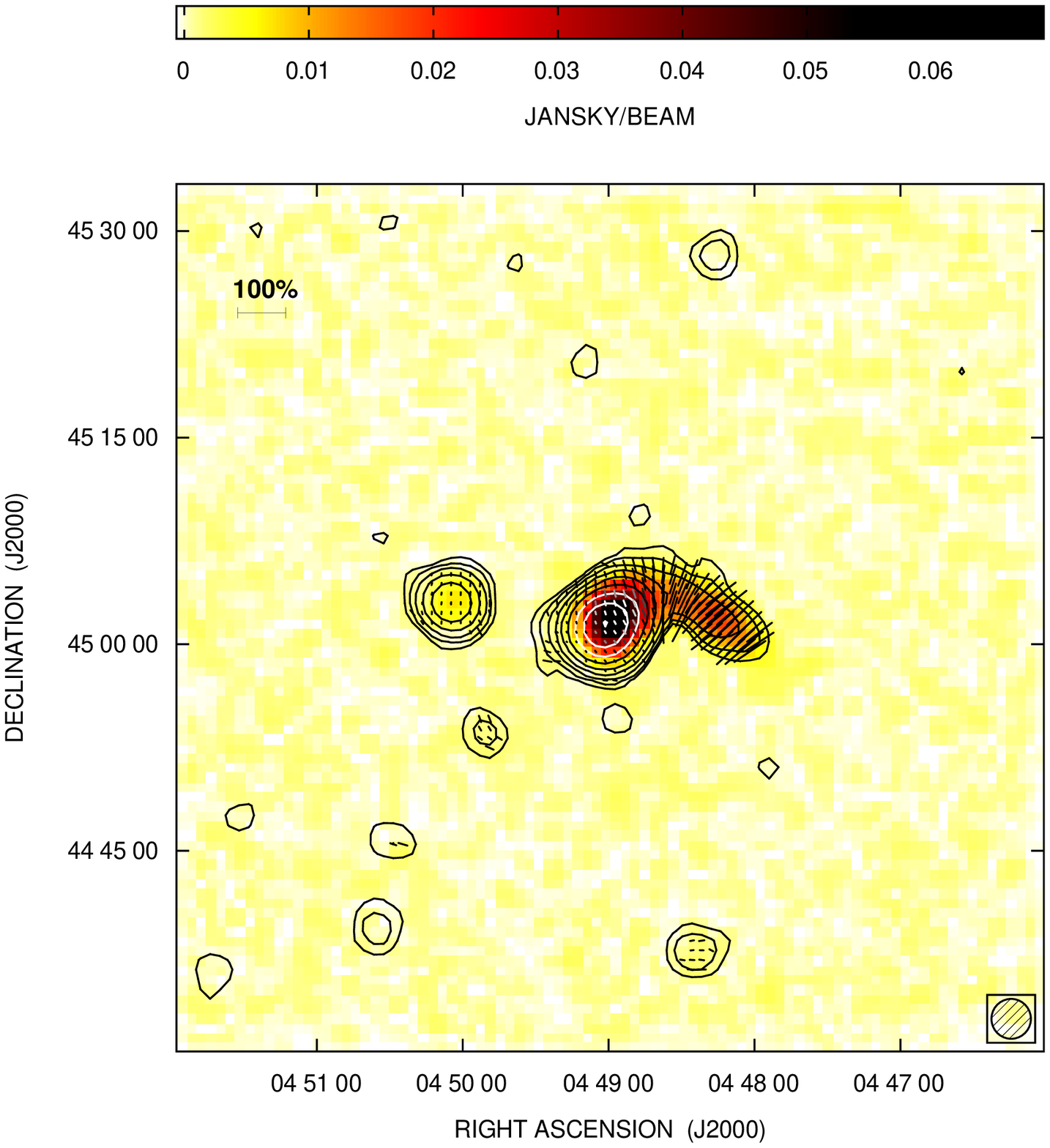}
\end{center}
\caption{SRT linearly-polarized intensity image of the galaxy cluster 3C\,129 resulting from the spectral average of the bandwidth between 6000 and 7200\,MHz. The polarized intensity is corrected for both the on-axis and the off-axis instrumental polarization. The noise level, after the correction for the polarization bias, is 0.6 mJy/beam. The FWHM beam is 2.9\arcmin, as indicated in the bottom-right corner. Contours refer to the CLEANed total intensity image. Levels start at 3$\sigma$ and increase by a factor of 2. The electric field polarization vectors were traced only for those pixels where the total intensity signal is above 5$\sigma$ and the error on the polarization angle is below 20\degree. The length of the vectors is proportional to the polarization percentage (with 100\% represented by the bar in the top-left corner), while their orientation represents the polarization angle. }
\label{3C129_field_pol}
\end{figure*}

The intensity, as measured in the OTF scans, is the convolution of the beam pattern with the true sky brightness distribution. An important point to stress is 
that the beam pattern of an alt-azimuth mount telescope rotates in the equatorial frame by the parallactic angle. Thus, when we combine scans taken
at different times, the orientation of the beam pattern will also be different, in general. This is illustrated in the right panel of Fig.\,\ref{beam} where we show the average beam pattern model re-projected into the equatorial RA-DEC frame and over imposed to the contour map of the 3C\,129 ``dirty map''. Note that the brightest part of the asymmetric first side-lobe of the beam is oriented by chance towards the tail of the radio galaxy.

SCUBE uses an elevation-dependent beam model for a proper deconvolution of the sky image from the antenna pattern. The deconvolution algorithm is based on 
a variant of the CLEAN procedure (H\"ogbom 1974) where the true sky brightness is represented by a collection of $\delta$-functions (also known as the CLEAN components).
It iteratively finds the peak in the image obtained from the SWT stacking of all images, and subtracts a fixed gain fraction of this point source flux (typically 0.1) convolved with the re-projected telescope ``dirty beam model'' from the individual images. In the re-projection, the exact elevation and parallactic angle for
 each pixel in the unstacked images are used. The residual images are stacked again and the CLEAN continues until a threshold condition is reached.
Given the low level of the beam side lobes compared to interferometric images, a shallow deconvolution is sufficient in our case, and we decided to stop the CLEAN at the first negative component encountered. As a final step, CLEAN components at the same position are merged, convolved with a circular Gaussian beam with FWHM 2.9\arcmin, and then restored back in the residuals to obtain a CLEANed image. We tried to CLEAN by using both the elevation-averaged beam model and the three elevation bins model. We found no significant differences in the CLEANed images. This is likely due to the fact that the 3C\,129 observations span a large range
of elevations. Therefore, in this work we decided to use the elevation-average beam model because of its higher signal-to-noise ratio.

The CLEANed image of the 3C\,129 field is shown in Fig.\,\ref{3C129_field_cleaned}. Thanks to the deconvolution of the beam pattern, we reached a dynamic range of about 800, and thus we can now observe the actual brightness and size of the tail of 3C\,129. 

\begin{figure*}

\begin{center}
\includegraphics[width=18cm]{fig11.ps}
\end{center}
\caption{SRT contour levels overlay the WENSS 325\,MHz image (left panel) and the VLA 1424\,MHz C+D image (right panel), see text. Levels start from 3$\sigma$ and increase by a factor of 2. Relevant sources are labeled and reported in Tab.\,2. The inset in the right panel shows a zoom of source C.}
\label{srt+wenss+vla}
\end{figure*}

\subsection{Polarization imaging}

The polarization imaging of Stokes parameters Q and U was performed following the same procedures described for the total intensity imaging: baseline subtraction, 
gridding, and SWT stacking. There are no critical dynamic range issues with the polarization image, and thus no deconvolution was
required. However, we corrected for the off-axis instrumental polarization by deconvolving the Stokes Q and U beam patterns shown in Fig.\,\ref{offaxispol}. In particular, 
SCUBE uses the CLEAN components derived from the deconvolution of the beam pattern from the total intensity image to subtract the spurious off-axis polarization from each 
individual Q and U scans before their stacking. The subtraction is performed by rotating the Q and U beam patterns by the appropriate parallactic angle for each pixel.

The resulting image of the linear polarization intensity from the radio sources in the field of the galaxy cluster 3C\,129 is presented in Fig.\,\ref{3C129_field_pol}.
We show in colors the polarized intensity ($P=\sqrt{Q^2+U^2}$) corrected for the bias introduced when combining the Stokes Q and U images, which we assumed 
to have equal noise Gaussian statistics with a measured standard deviation $\sigma_{\rm Q,\,U}=0.6$ mJy/beam (see Appendix B in Killeen et
 al. 1986). The peak polarized intensity is 70 mJy/beam.

In Fig.\,\ref{3C129_field_pol}, we overlaid the total intensity contour levels and the polarization vectors to the image of the polarized intensity.
The length of the vectors is proportional to the fractional polarization, while their orientation gives the polarization angle. They have been plotted only in those
regions where the error in the polarization angle is less than 20\degree~and the total intensity is above 5 $\sigma_{\rm I}$.

Polarization is detected for both 3C\,129 and 3C\,129.1 with a global fractional polarization of 0.26 and 0.07, respectively.
In 3C\,129, the polarization angle of the radiation electric field (not corrected for the Faraday rotation effect) is perpendicular to the tail along its entire length.
The fractional polarization increases from 0.05 in the head of the radio source up to 0.67 at the end of the tail.
A detailed analysis of the polarization properties is presented in Sect.\,\ref{polanal}.

\section{Spectral analysis}
\label{specanal}

We analyzed the spectral properties of the radio sources in the 3C\,129 field by complementing and comparing the SRT image with other images available in the literature
at lower frequencies. In particular, we compared the SRT image at 6600\,MHz with the VLA Low-Frequency Sky Survey redux (VLSSr; Lane et al. 2014),
the Westerbork Northern Sky Survey (WENSS; Rengelink et al. 1997), and with the NRAO VLA Sky Survey (NVSS; Condon et al. 1998). 

In the left panel of Fig.\,\ref{srt+wenss+vla}, we show the SRT contour levels overlaying the WENSS image at 325\,MHz, which 
has a noise level of 5 mJy/beam and a resolution 
of $76\arcsec\times 54\arcsec$~FWHM beam. In addition to 3C\,129 and 3C\,129.1, some fainter sources were detected by the SRT. In order to distinguish genuine detections from spurious noise spikes, we searched for a counterpart in the WENSS and in the NVSS images  at the positions of these putative sources. 
We found that most of these sources, labeled with letters A-N in Fig.\,\ref{srt+wenss+vla}, have a counterpart in the WENSS and NVSS images except for source C. Furthermore, there are two sources, labeled J and K, which are visible in the WENSS and the NVSS at a level $> 10\sigma$ but were not detected at the sensitivity level of the SRT image.

\begin{figure*}

\begin{center}
\includegraphics[width=18cm]{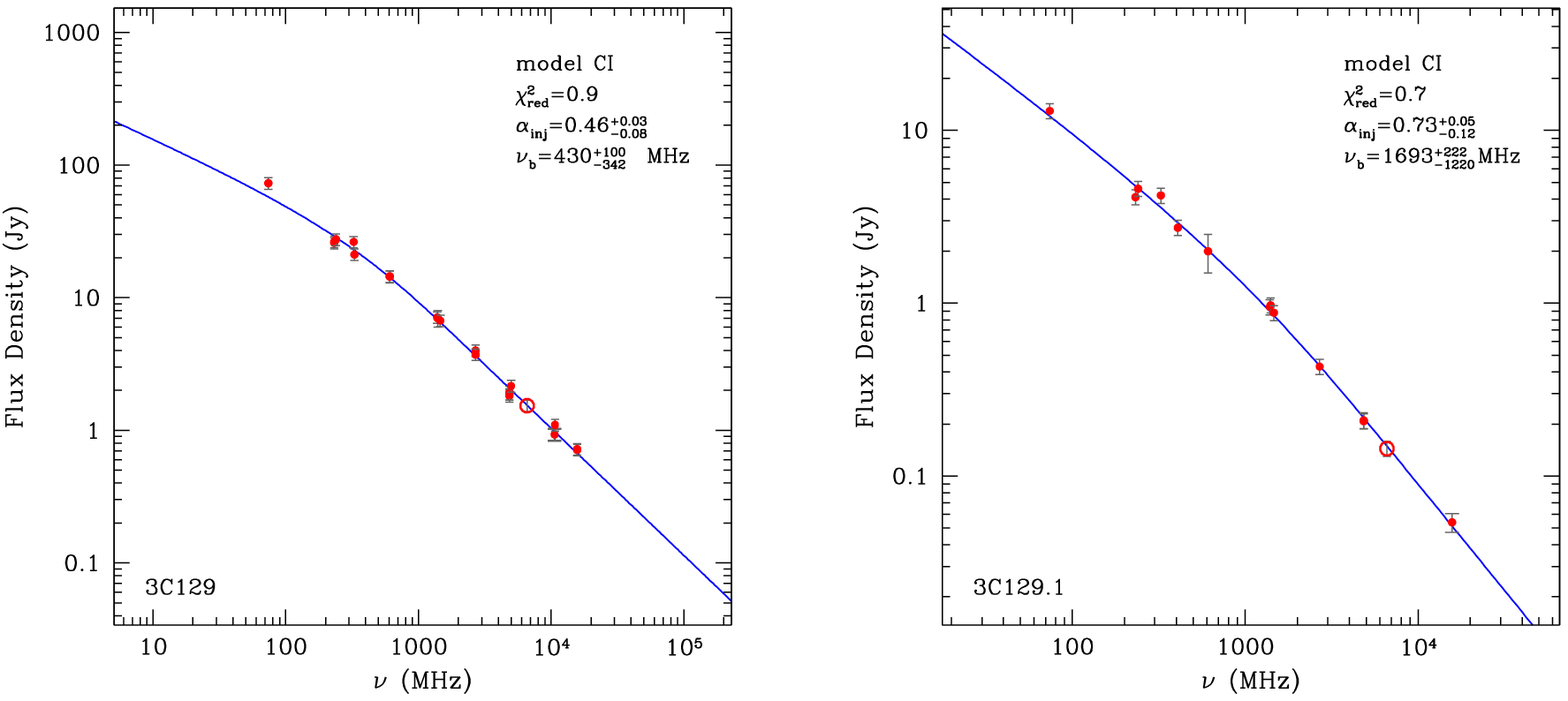}
\end{center}
\caption{Integrated spectra of 3C\,129 (left) and 3C\,129.1 (right). The SRT flux density measurements, indicated by the open dot symbols, are in perfect agreement with the
 data at close frequencies taken from the literature. The error bars of the SRT flux density measurements are comparable with the size of the symbols. 
The continuous line is the best fit for the CI model, see text.}
\label{3C129_spectra}
\end{figure*}

From the NRAO archive, we recovered the L-band programs AH746 and AH744. The observations were taken with the VLA in C and D configurations, respectively, using for both a bandwidth of 50\,MHz at two IFs of 1385\,MHz and 1465\,MHz. The data sets were calibrated independently with the Astronomical Image Processing System (AIPS) package following the standard procedures and then combined together with task DBCON to improve uv-coverage and sensitivity. In the right panel of Fig.\,\ref{srt+wenss+vla}, we show the SRT contour levels overlaying the VLA C+D image. We average the two IFs, obtaining an image with a frequency of 1424\,MHz, a noise level of 0.27 mJy/beam, and a FWHM beam resolution of 19\arcsec$\times$14\arcsec~with PA=$-15$\degree. The comparatively higher resolution of the VLA image with respect to that of the WENSS and the SRT images, makes it possible to appreciate the morphology of 3C\,129 in better detail. The twin jets emerging from the AGN hosted in the head of the source are bent back by the ram pressure exerted by the intracluster medium
and appear to merge forming the 500\,kpc long tail.  The SRT contours show an elongation of the head of the source toward the direction of 3C\,129.1. 
If real, this feature appears to be roughly aligned with an extension of the famous  ``crosspiece'' (see e.g. Fig.\,2 in Lane et al. 2002) marginally visible in the VLA image, 
but the association is not obvious given the rather different resolution of the two observations. Apart from 3C\,129 and 3C\,129.1, several faint mJy level point sources are visible in the VLA image. One of these appears to be a small double source which is possibly associated with the source labeled C, see the inset in the right panel of Fig.\,\ref{srt+wenss+vla}. Another faint point source is visible close to the tail, just half-way from the head.

\begin{table*}
\label{Observations}
\caption{SRT flux density measurements at 6600 MHz. Uncertainties are at $1\sigma$ level. Coordinates refer to the intensity peak in the SRT image.}             
\centering          
\begin{tabular}{ c c c c c c c c}     
\hline\hline
Source         &  Label     & RA J2000       & DEC J2000                & $S_{\nu}$           & $\alpha_{\rm 1400\,MHz}^{\rm 325\,MHz}$   & $\alpha_{\rm 6600\,MHz}^{\rm 1400\,MHz}$ & SPC                 \\	  
               &            & (h m s)        &  (d \arcmin~\arcsec)     & (mJy)               &       &      &                     \\
\hline\hline
3C\,129             &  -         &  04 49 03      &   45 01 44          & $ 1534 \pm 6   $    & $0.91 \pm 0.10  $ & $0.98  \pm   0.09  $  &   0.07 \\ 
3C\,129.1           &  -         &  04 50 06      &   45 03 06          & $ 144  \pm 3   $    & $1.00 \pm 0.01  $ & $1.23  \pm   0.01  $  &   0.23  \\ 
NVSS J044818+452818 &  A         &  04 48 16      &   45 28 25          & $ 12.7 \pm 0.9 $    & $0.40 \pm 0.06  $ & $0.41  \pm   0.05  $  &   0.01   \\ 
NVSS J044908+452009 &  B         &  04 49 11      &   45 20 33          & $ 7.2  \pm 1.7 $    & $0.66 \pm 0.07  $ & $0.76  \pm   0.15  $  &   0.10    \\ 
  -                 &  C         &  04 48 58      &   44 54 44          & $ 7.5  \pm 1.6 $    & -                 & $-0.41 \pm   0.17  $  &  -    \\ 
NVSS J044954+445330 &  D         &  04 49 52      &   44 53 48          & $ 11.5 \pm 1.3 $    & $0.61 \pm 0.08  $ & $0.43  \pm   0.07  $  &  -0.18    \\ 
NVSS J045130+442928 &  E         &  04 51 29      &   44 27 46          & $ 5.8  \pm 0.8 $    & $1.00 \pm 0.04  $ & $0.96  \pm   0.09  $  &  -0.04    \\ 
NVSS J045032+444528 &  F         &  04 50 30      &   44 46 04          & $ 10.1 \pm 1.4 $    & $0.84 \pm 0.04  $ & $0.75  \pm   0.09  $  &  -0.09    \\ 
NVSS J045036+443920 &  G         &  04 50 35      &   44 39 38          & $ 14.5 \pm 1.8 $    & $0.51 \pm 0.08  $ & $0.16  \pm   0.08  $  &  -0.34   \\ 
NVSS J045145+443603 &  H         &  04 51 42      &   44 36 05          & $ 8.4  \pm 1.6 $    & $0.68 \pm 0.04  $ & $1.02  \pm   0.12  $  &   0.34    \\ 
NVSS J044827+443753 &  I         &  04 48 26      &   44 37 59          & $ 17.7 \pm 1.3  $   & $0.78 \pm 0.02  $ & $0.80  \pm   0.05  $  &   0.02    \\ 
NVSS J044703+452002 &  J         &     -          &      -              &    $<4.5$           & $1.36 \pm 0.06  $ & $>0.67             $  &     -   \\ 
NVSS J044640+445220 &  K         &     -          &      -              &    $<4.5$           & $1.31 \pm 0.06  $ & $>0.33             $  &     -   \\ 
NVSS J044756+445025 &  L         &  04 47 56      &   44 51 11          & $ 5.4 \pm 1.8 $     & $0.61 \pm 0.18  $ & $0.58  \pm   0.21  $  &  -0.02    \\  
NVSS J044941+452713 &  M         &  04 49 39      &   45 27 38          & $ 4.4 \pm 1.1 $     & $0.88 \pm 0.11  $ & $0.63  \pm   0.16  $  &  -0.25    \\  
NVSS J045035+450744 &  N         &  04 50 34      &   45 08 08          & $ 4.1 \pm 1.6 $     &  -                & $-0.23 \pm   0.28  $  &  -    \\  
\hline\hline
\multicolumn{8}{l}{\scriptsize Col.1: cross-identification; Col.2: source label (see Fig.\,\ref{srt+wenss+vla}); Cols.3 and 4: coordinates of the peak in the SRT image; Col.5: flux density at 6600\,MHz;}\\
\multicolumn{8}{l}{\scriptsize Col.6: spectral index between 325 and 1400\,MHz; Col.7: spectral index between 1400 and 6600\,MHz; Col.8: spectral curvature $SPC=\alpha_{\rm 6600\,MHz}^{\rm 1400\,MHz}-\alpha_{\rm 1400\,MHz}^{\rm 325\,MHz}$.}\\
\end{tabular}   
\end{table*} 

In Tab.\,2, we listed the basic properties for the radio sources in the field of the SRT image. We calculated the flux density for 3C\,129 and 3C\,129.1 by integrating 
the total intensity down to the 3$\sigma$ isophote. For all the other sources in the field, which were point-like or only slightly resolved, we calculated the flux density by means of a bi-dimensional Gaussian elliptical model characterized by six free parameters: peak position, peak amplitude, $\rm FWHM_{\rm maj}$, $\rm FWHM_{\rm min}$, and position angle. In Tab.\,2, we reported the integrated model flux density corresponding to the best fit parameters. We tested the reliability of the fitting procedure by injecting in the SRT image fake sources of progressively lower intensity and we checked 
the consistency of the input and output flux densities. Sources L, M and N, with peak intensities of respectively $5.6\pm 1.5$, $5.3\pm\ 1.5$, and $5.2\pm 1.5$ mJy/beam at 6600\,MHz, are the faintest sources listed in Tab.\,2 and the quoted flux densities should be considered with care. However, by assuming they are point-like sources, the estimated flux densities were consistent within the uncertainties with the peak values measured in the SRT image.

Along with the SRT coordinates and flux densities of the sources, we also report their NVSS name, the global spectral indices ($S_{\nu}\propto \nu^{-\alpha}$) between 325 and 1400\,MHz, between 1400 and 6600\,MHz, and the spectral curvature (SPC= $\alpha_{\rm 6600\,MHz}^{\rm 1400\,MHz}-\alpha_{\rm 1400\,MHz}^{\rm 325\,MHz}$; Murgia et al. 2011). We note that sources J and K have a steep radio spectrum, which may explain why they have not been detected in the SRT image.

\subsection{Integrated spectra}

In order to check the accuracy of the flux scale calibration in our SRT observations, we collected all of the spectral information available in the literature for the
 radio galaxies 3C\,129 and 3C\,129.1. We made use of the on-line Astrophysical CATalogs support System (CATS; Verkhodanov et al. 1997) and the
Nasa Extragalactic Database (NED) to recover data from catalogs at different frequencies. The compilation also includes the GMRT measurements at 240 and 610\,MHz by Lal \& Hao (2004). 

The integrated spectra of 3C\,129 and 3C\,129.1 are shown in Fig.\,\ref{3C129_spectra}. For both spectra, we found that the SRT flux density measurements, indicated by the open dot symbols, are
in perfect agreement with the data at close frequencies taken from the literature.

The morphology, the spectral indices, and the spectral curvatures of both 3C\,129 and 3C\,129.1, indicate that these are currently active sources whose synchrotron emission is dominated 
by the electron populations with GeV energies injected by the radio jets and accumulated during their entire lives (Murgia et al. 2011). Therefore, 
following Murgia et al. (1999), we fitted the integrated spectra with the continuous injection model (CI; Pacholczyk 1970). The CI model is characterized by three 
free parameters: the injection spectral index ($\alpha_{\rm inj}$), the break frequency ($\nu_{b}$), and the flux normalization.
In the context of the CI model, it is assumed that the spectral break is due to the energy losses of the relativistic electrons. 
For high-energy electrons, the energy losses are primarily due to the synchrotron radiation itself and to the inverse Compton scattering of the Cosmic Microwave 
Background (CMB) photons. According to the synchrotron theory (e.g. Blumenthal \& Gould 1970, Rybicki \& Lightman 1979), it is possible to relate the break frequency to the time elapsed since
the start of the injection:

\begin{equation}
t_{\rm syn}= 1590 \frac{B^{0.5}}{(B^2+B_{\rm IC}^2) [(1+z)\nu_{\rm b}]^{0.5}}~\rm Myr,
\label{synage}
\end{equation}
 
where $B$ and $B_{\rm IC}=3.25(1+z)^2$ are the source magnetic field and the inverse Compton equivalent magnetic field associated with the CMB, respectively, and an isotropic distribution of the electrons pitch angles was assumed.

Expansion losses may play a role for rapidly expanding plasma on a timescale given by

\begin{equation}
t_{\rm exp} \sim  \left(\frac{R}{\rm kpc}\right) \left(\frac{v_{\rm exp}}{1000\, \rm km/s}\right)^{-1} ~\rm Myr,
\label{explosses}
\end{equation}
 
where $R$ and $v_{\rm exp}$ are the characteristic size and expansion speed, respectively.

While radiative losses affect the shape of the high-frequency spectrum, expansion losses also affect the normalization of the low-frequency
spectrum (see Murgia et al. 1999 for a more detailed discussion of the combined effects in the modeling of the integrated spectrum).
  
During the active phase, the evolution of the integrated spectrum is determined by the shift with time of $\nu_{\rm b}$ to lower and lower frequencies. Indeed, the spectral
break can be considered to be a clock indicating the time elapsed since the injection of the first electron population. Below and above $\nu_{\rm b}$, the spectral indices are
respectively $\alpha_{\rm inj}$ and $\alpha_{\rm inj}$+0.5.

For 3C\,129 the best fit of the CI model to the observed radio spectrum yields a break frequency of $\nu_{\rm b}=430\pm 220$\,MHz with $\alpha_{\rm inj}=0.46\pm 0.06$, while for 3C\,129.1 the best fit parameters for the CI model are  $\nu_{\rm b}=1693\pm721$\,MHz and $\alpha_{\rm inj}=0.73\pm 0.08$. The uncertainty on the break frequency derived from the fit is quite large ($\simeq 40$\%) due to the smooth curvature of the CI model. However, a more precise estimate can be obtained from the fit of the local spectra presented in the following sub-section.

\begin{figure*}

\begin{center}
\includegraphics[width=18cm]{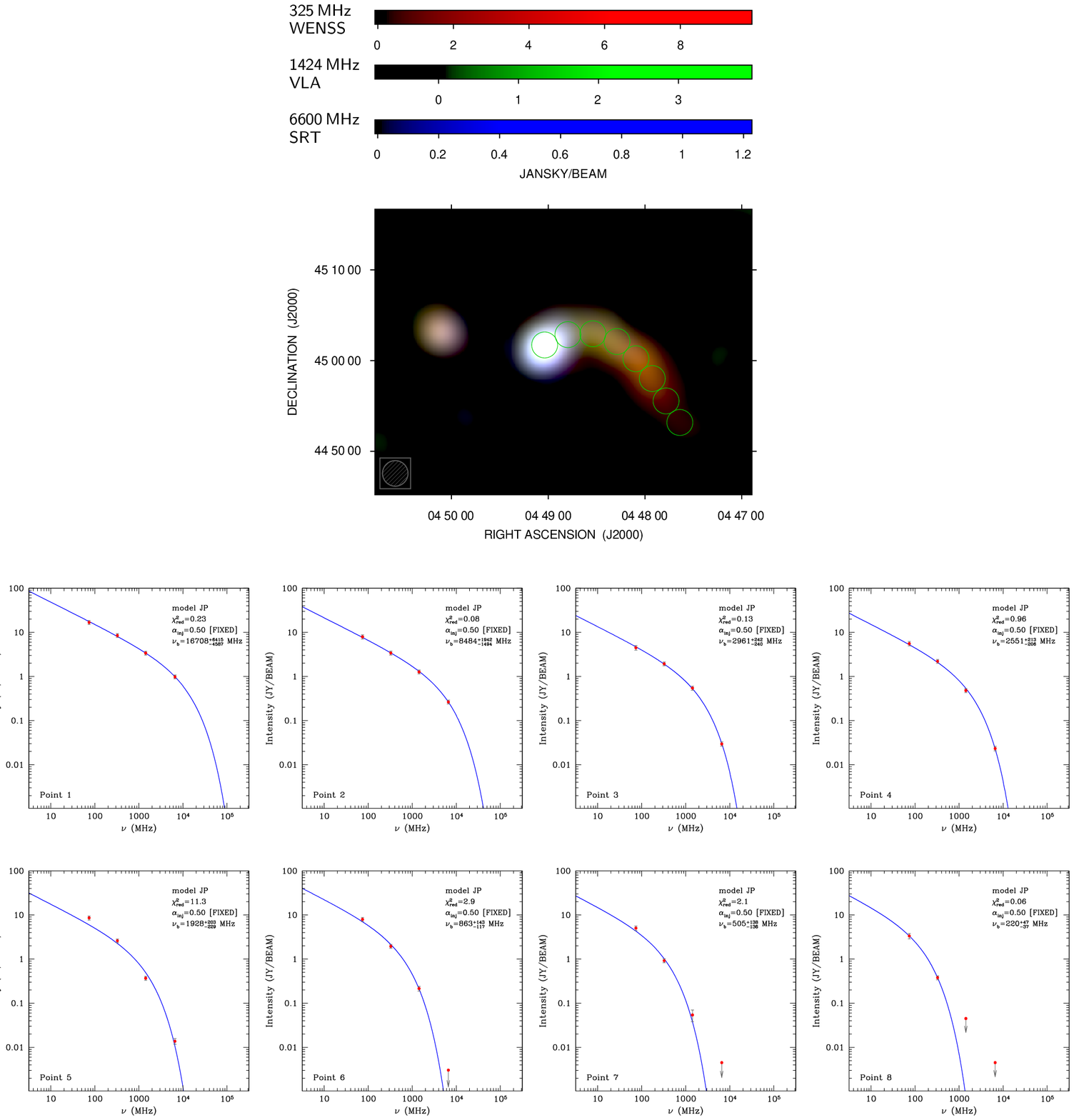}
\end{center}
\caption{Top panel: three color composite images of 3C\,129 obtained from WENSS (red), VLA (green), and SRT (blue) images all convolved to a common resolution of 2.9\arcmin. The color turns smoothly from bluish to redish, going from the head to the end of the tail as a result of the intense radiative cooling of the high-energy synchrotron electrons. Middle and bottom panels: synchrotron spectra sampled at eight different locations along the head and the tail of 3C\,129, as shown by the circular regions in the top panel. The lines represent the best fit of the JP model.}
\label{3C129_local_fit}
\end{figure*}

\subsection{Local spectra}

The integrated spectra presented in Fig.\,\ref{3C129_spectra} are characterized by a remarkable frequency coverage, but completely lack in spatial resolution. 
On the other hand, by combining the SRT image at 6600 MHz with the VLSSr at 74\,MHz, the WENSS at 325\,MHz, and the VLA C+D image at 1424\,MHz, it is possible to analyze the variation
of the synchrotron spectrum along the tail of 3C\,129, albeit using a limited number of frequencies. We convolved the VLSSr, WENSS, and the VLA C+D array images at the same resolution
as that of the SRT image, and we extracted the spectra in eight circular regions along the source, as shown in Fig.\,\ref{3C129_local_fit}. The regions are four pixels in diameter so that 
the spectra are sampled at effectively independent locations. The last box at the end of the tail is located at about 19.6\arcmin~from the host galaxy, which corresponds to a physical distance of 488\,kpc assuming that the source is on the plane of the sky. We then fitted the observed spectra with the JP model (Jaffe \& Perola 1973), which also has three free parameters
  $\alpha_{\rm inj}$, $\nu_{b}$, and normalization like the CI model. The JP model however describes the spectral shape of an isolated electron population injected at a specific instant
in time with an initial power law energy spectrum with index $p=2\alpha_{\rm inj}+1$. Radiative losses deplete the high-energy electrons, and the emission spectrum develops an
exponential cut-off beyond the break $\nu_{b}$, which progressively shifts to lower frequency with time according to Eq.\,\ref{synage}.

\begin{figure*}
\begin{center}
\includegraphics[width=18cm]{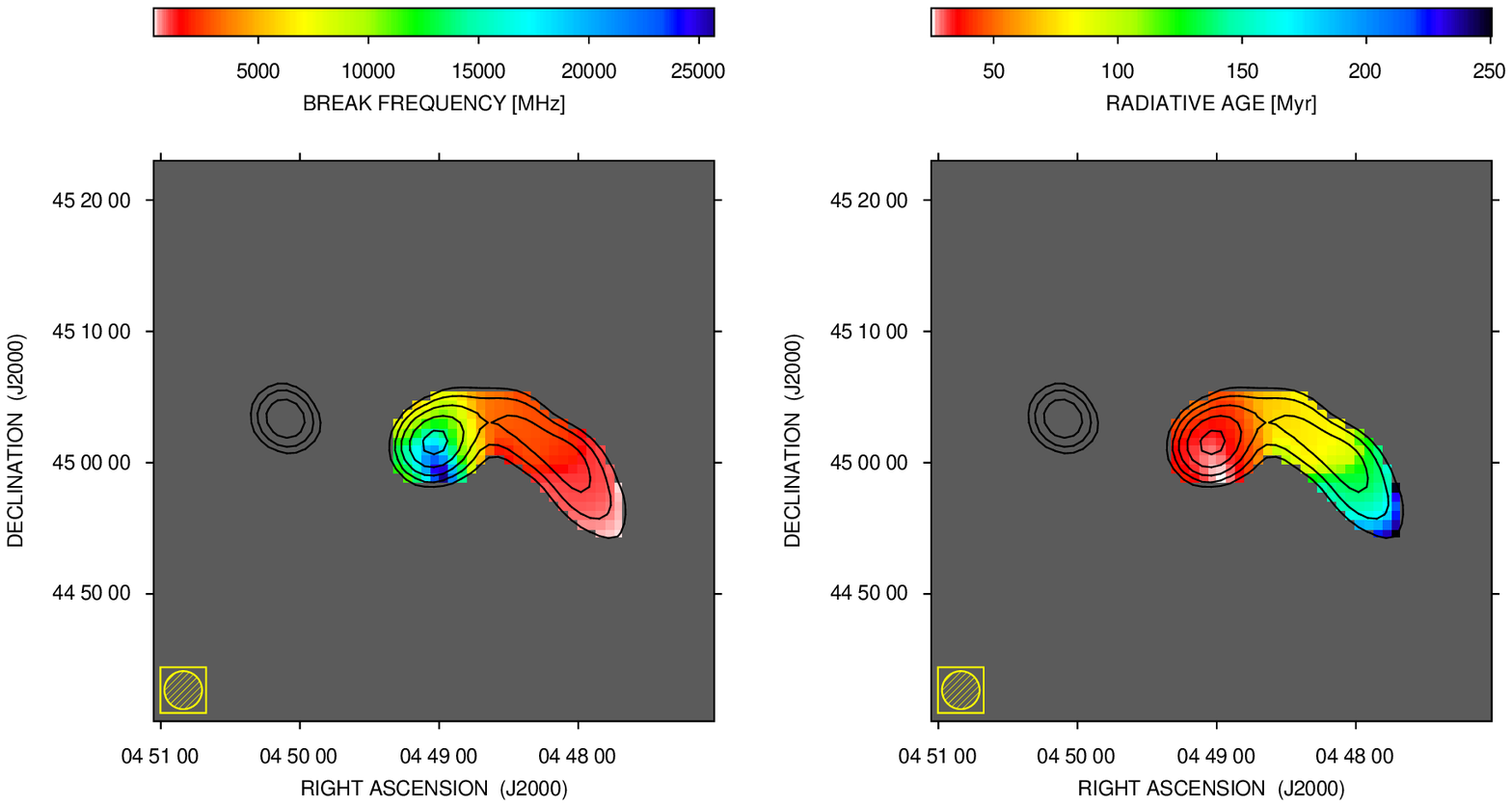}
\end{center}
\caption{Maps of break frequency (left) and radiative age (right) for 3C\,129.}
\label{3C129_age_map}
\end{figure*}

To limit the number of free parameters, we fixed $\alpha_{\rm inj}=0.5$. This value provides a good fit to the spectrum of the head and is in agreement with the fit of the CI model to the integrated spectrum of 3C\,129. The measured break frequency decreases systematically along the tail in agreement with the scenario in which the oldest particles are those at larger distance from the AGN. The minimum break frequency measured at the end of the tail is $\nu_{b}=220\pm 42$\,MHz, in agreement, within the error, with the spectral break measured in the integrated spectrum. In a previous spectral analysis, Feretti et al. (1998) found a break frequency of 1900\,MHz at a distance of 14\arcmin~from the head of 3C\,129, which is the maximum distance
allowed by the sensitivity of their images. This distance lies in between regions number four and five in our analysis. At these locations, we measured break frequencies of $\nu_{b}=2551\pm 209$\,MHz and  $\nu_{b}=1928\pm216$\,MHz.

The value of the minimum break frequency which we directly measured at the end of the tail is also fully consistent with the value of 240\,MHz suggested
 by Lal \& Rao (2004), based on their considerations on GMRT observations.

\subsection{3C\,129 radiative age and galaxy velocity}

We derived the radiative age from Eq.\,\ref{synage} using for the magnetic field strength the minimum energy value. The minimum energy magnetic field was
calculated assuming, for the electron energy spectrum, a power law with index $p=2\alpha_{\rm inj}+1=2$ and a low energy cut-off at a Lorentz factor $\gamma_{\rm low}=100$.
In addition, we assumed that non-radiating relativistic ions have the same energy density as the relativistic electrons and, in the specific case of 3C\,129, 
we considered an ordered magnetic field oriented at 90\degree~to the line-of-sight (see Sect.\,\ref{polanal}). With this choice, the minimum energy field is 12\% lower than in the case of a totally random field (Murgia et al. 2012). We used the luminosity at 74\,MHz, since radiative losses are less important at low frequencies. It is found that the minimum energy field, at the locations of the same eight regions used to extract the local spectra, varies
from 4\,$\mu$G close to the head, down to 2.8\,$\mu$G at the end of the tail. The average minimum field strength is $\langle B_{\rm min}\rangle\simeq 3.4 \,\mu$G. Our values for the minimum energy field fit in between those calculated by Feretti et al. (1998) and Lal \& Rao (2004).

Assuming for the source's magnetic field $\langle B_{\rm min}\rangle=3.4 \mu G$ and for the break frequency the lowest measured value of $\nu_{b}=220\pm 42$\,MHz, the radiative age of 3C\,129 is of the order of $t_{\rm syn}\simeq 267\pm 26$ Myrs. By assuming a linear projected length of about 488\,kpc for the tail, this implies an average advancing velocity for the galaxy of $v_{\rm gal}\simeq 1785\pm 175\cdot \sec(i)$\,km/s (where $i$ is the inclination of the tail with respect to the plane of the sky). According to Lane et al. (2002), if 3C\,129 is assumed to be a member of an infalling galaxy cluster, then the infalling velocity of matter onto the 3C\,129 cluster is expected to be $v_{\rm inf}=\sqrt{3k T_{\rm ICM}/\mu}$, where $\mu$ is the mean molecular mass of the gas. Given the value of $kT_{\rm ICM} = 5.5\pm 0.2$\,keV for the intra-cluster medium temperature (Leahy \& Yin 2000), the infall velocity is expected to be 1630 km/s (Lane et al. 2002), i.e. comparable within the uncertainty with the advancing speed obtained from the spectral analysis. Given the sound speed of $c_{\rm s}=\sqrt{(5/3)k T_{\rm ICM}/\mu}=1217\pm 22$ km/s reported by Lane et al. (2002), we deduced that 3C\,129 was moving supersonically with a Mach number as high as $M=v_{\rm gal}/c_{\rm s}=1.47$, which is very close to the speed invoked by those authors to explain the Mach cone opening angle of $\sim 90$\degree~for the shock-wave front that would be associated with the ``crosspiece'' (see Fig.\,2 in that paper).

In Fig.\,\ref{3C129_age_map}, we show the break frequency map and the radiative age map of 3C\,129 obtained by fitting the JP model pixel-by-pixel.

\section{Polarization analysis}
\label{polanal}

%\subsection{Fractional polarization trend}

As outlined in Sect.\,4.3, we determined the linear polarization in 3C\,129, 3C\,129.1 and, tentatively, in the field point sources named  D, I, and F. Here we focus on the analysis of the polarization of 3C\,129. The degree of fractional polarization in 3C\,129 at 6600\,MHz increases from values of 5\% close to the head up to values as high as 70\% far down the tail. The polarization angle of the electric vector is perpendicular to the source's ridge line, indicating that the magnetic field is highly ordered and aligned with the tail (see Fig.\,\ref{3C129_field_pol}).

\begin{figure*}

\begin{center}
\includegraphics[width=18cm]{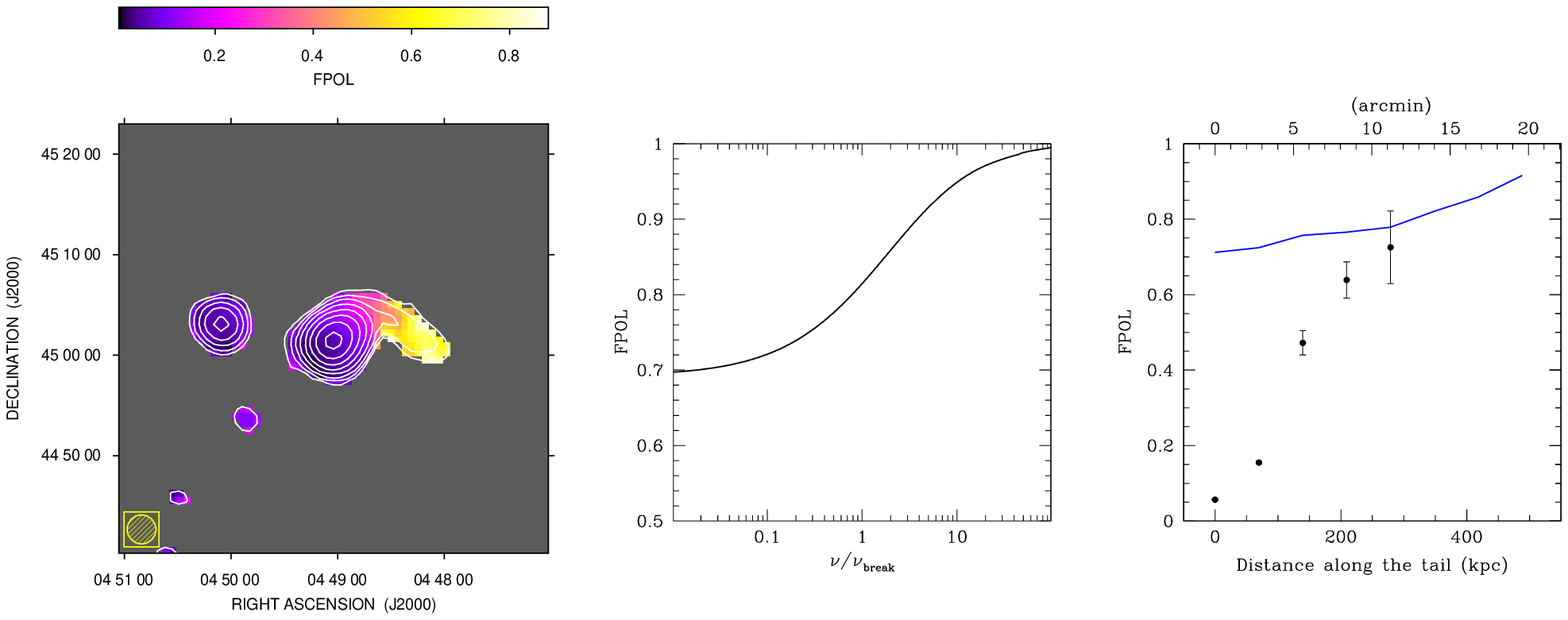}
\end{center}
\caption{Left: image of 3C\,129 fractional polarization at 6600\,MHz measured with the SRT. Middle: theoretical fractional polarization level at $\nu=6600$\,MHz as a function of synchrotron aging for $\alpha_{\rm inj}=0.5$. Right: profile of fractional polarization measured in five independent circular regions (these are five out of the eight regions shown in Fig.\,\ref{3C129_local_fit} top panel), from the head to the end of the tail of the source. The blue line is the theoretical value expected for a totally ordered source's magnetic field oriented at 90\degree~with respect to the line-of-sight, which we traced taking into account the theoretical expectation shown in the Middle panel and the break frequency trend determined from the spectral fit of the JP model.}
\label{3C129_field_order}
\end{figure*}

In the left panel of Fig.\,\ref{3C129_field_order}, we show the image of 3C\,129 fractional polarization. The synchrotron radiation theory predicts that the intrinsic fractional polarization level increases as the emission
spectrum steepens due to radiative losses. In the middle panel of Fig.\,\ref{3C129_field_order}, we traced the theoretical fractional polarization level expected for $\alpha_{\rm inj}=0.5$ as a function of the ratio of the observing frequency to the break frequency for a totally ordered magnetic field oriented at 90\degree~with respect to the line-of-sight. In the right panel of Fig.\,\ref{3C129_field_order}, we plotted the observed fractional polarization level  measured in five independent circular regions from the head to the end of the tail of the source. The blue line represents the expected level traced by assuming, for the break frequency, the value obtained from the spectral fits of the JP model derived in Sect.\,5. 

In general, the observed fractional polarization is reduced with respect to the theoretical value by the combination of several effects: i) the magnetic field is intrinsically disordered inside the radio source; ii) the radio source has a complex and unresolved structure on scales smaller than the linear size of the resolution of the observation (which is about 70\,kpc in the case of the SRT images);  
 iii) there is a significant Faraday rotation occurring internally to the synchrotron emitting plasma which causes an unrecoverable signal depolarization inside the finite frequency bandwidth
 of our observing channels; and iv) the polarization angle is modified by an external Faraday screen whose RM auto-correlation length is smaller than the observing beam (see e.g. Burn 1966, Tribble 1992, Murgia et al. 2004, Laing et al. 2008, Vacca et al. 2012).

The first two are essentially geometric effects, while the last two effects are most relevant at low frequencies due to the $\lambda^2$ dependence of the Faraday rotation.
If we suppose that Faraday rotation effects iii) and iv) are negligible at 6600\,MHz, the observed fraction polarization can be related to the theoretical one by the expression

\begin{equation}
\rm FPOL_{\rm obs}=\rm FPOL_{\rm th}~\frac{B_{\rm ord}^2}{B_{\rm ord}^2+B_{\rm rnd}^2}
\end{equation}

where $B_{\rm ord}$ and $B_{\rm rnd}$ are the ordered and random field components, respectively (see Feretti et al. 1998). Close to the source's head, the observed fractional polarization is 
much lower than the theoretical one, indicating a prevailing disordered magnetic field. Down the tail, 
at 10\arcmin~from the head, the observed fractional polarization approaches the theoretical value, indicating that the ordered magnetic field accounts for 93\% of the total field strength.

\section{Summary}

We have presented a wide-band spectral-polarimeric study of the galaxy cluster 3C\,129 performed with the new SARDARA backend installed at the 
SRT. The results are summarized in the following.

Total intensity images at an angular resolution of about 2.9\arcmin~were obtained for the tailed radio galaxy 3C\,129 and for 
13 more sources in the field, including 3C\,129.1 at the galaxy center. We modeled the total intensity telescope 
beam pattern and the off-axis instrumental polarization patterns.
We used these models to deconvolve the telescope beam from the total intensity 
images and to correct the polarized images for the off-axis instrumental polarization.

The SRT data were used, in combination with literature data at
lower frequencies, to derive the variation of the synchrotron spectrum of 3C\,129 along the tail of the radio source. 
We computed the global spectra of 3C\,129 and 3C\,129.1 that we used to compare the SRT total intensity measurements with the
literature data at other frequencies. We found that the new SRT data points are in remarkable agreement with both data
from the literature at close frequencies and the best model fit to the global synchrotron spectrum. This result confirms the
accuracy of the flux density scale calibration at the SRT.
In addition, we found that the radio spectrum in 3C\,129 progressively steepens with distance from the core,
and at each location it is steeper at higher frequencies. The local spectra are well fitted by models
involving synchrotron energy losses, assuming a continuous isotropization of the pitch-angle distribution
(JP model). The break frequency obtained
by spectral fits decreases with increasing distance from
the host galaxy. Assuming that the magnetic field is at the equipartition
value, we showed that the lifetimes of radiating electrons after
synchrotron and Inverse Compton losses result in a radiative age for 3C\,129 of the order of $t_{\rm syn}\simeq 267\pm 26$ Myrs. 
By assuming a linear projected length of 488\,kpc for the tail, this implies a relative average advancing velocity for the galaxy of 
$v_{\rm gal}\simeq 1785\pm 175\cdot \sec(i)$\,km/s (where $i$ is the inclination of the tail with respect to the plane of the sky).  Given the sound speed of $c_{\rm s}=1217\pm 22$  km/s, we deduced that 3C\,129 is moving 
supersonically with a Mach number as high as $M=v_{\rm gal}/c_{\rm s}=1.47$, which is very close to the speed invoked by Lane et al. (2002) to explain the Mach cone 
opening angle of $\sim 90$\degree~for the shock-wave front that would be associated with the ``crosspiece'' observed in front of the host galaxy.

Linearly polarized emission is clearly detected for both 3C\,129 and 3C\,129.1. The linear polarization determined for 3C\,129 with the SRT at 6600\,MHz reaches levels as high as 70\%
in the faintest region of the source where the magnetic field is aligned with the direction of the tail.
This result can be interpreted as a combination of the spectral steepening due to the radiative losses of the synchrotron electrons, 
leading to an intrinsically higher fractional polarization, and to the ordering of the magnetic field that increases with the increase of the distance from the head.

\section*{Acknowledgements}
We are sincerely grateful to the anonymous referee for her/his comments and suggestions which helped us to improve this paper.
The Sardinia Radio Telescope (SRT, Bolli et al. 2015, Prandoni et al. in prep.)
is funded by the Department of University and Research (MIUR),
Italian Space Agency (ASI), and the Autonomous Region of Sardinia (RAS) 
and is operated as National Facility by the National Institute for Astrophysics (INAF).
The development of the SARDARA backend has been funded by the Autonomous Region of Sardinia (RAS)
using resources from the Regional Law 7/2007 "Promotion of the scientific research and
technological innovation in Sardinia" in the context of the research project CRP-49231 (year
2011, PI Possenti): "High resolution sampling of the Universe in the radio band: an unprecedented
instrument to understand the fundamental laws of the nature". F. Loi gratefully acknowledges Sardinia Regional Government for the
financial support of her PhD scholarship (P.O.R. Sardegna F.S.E. Operational Programme of the Autonomous Region of Sardinia, European
Social Fund 2007-2013 - Axis IV Human Resources, Objective l.3, Line of Activity l.3.1.). This research was partially supported by PRIN-INAF2014.
The National Radio Astronomy Observatory is operated by Associated Universities, Inc., under contract with the National Science Foundation. This research made use of the NASA/IPAC Extragalactic Database (NED) which is operated by the Jet Propulsion Laboratory, California Institute of Technology, under contract with the National Aeronautics and Space Administration. This research made use also of the CATS Database (Astrophysical CATalogs support System).

%%%%%%%%%%%%%%%%%%%%%%%%%%%%%%%%%%%%%%%%%%%%%%%%%%

%%%%%%%%%%%%%%%%%%%% REFERENCES %%%%%%%%%%%%%%%%%%

% The best way to enter references is to use BibTeX:

%\bibliographystyle{mnras}
%\bibliography{example} % if your bibtex file is called example.bib

% Alternatively you could enter them by hand, like this:
% This method is tedious and prone to error if you have lots of references

%%%%%%%%%%%%%%%%%%%%%%%%%%%%%%%%%%%%%%%%%%%%%%%%%%

%%%%%%%%%%%%%%%%% APPENDICES %%%%%%%%%%%%%%%%%%%%%

\appendix
\section{Image stacking and de-stripping by mixing of wavelet transform coefficients}

\begin{figure*}
\begin{center}
\includegraphics[width=14cm]{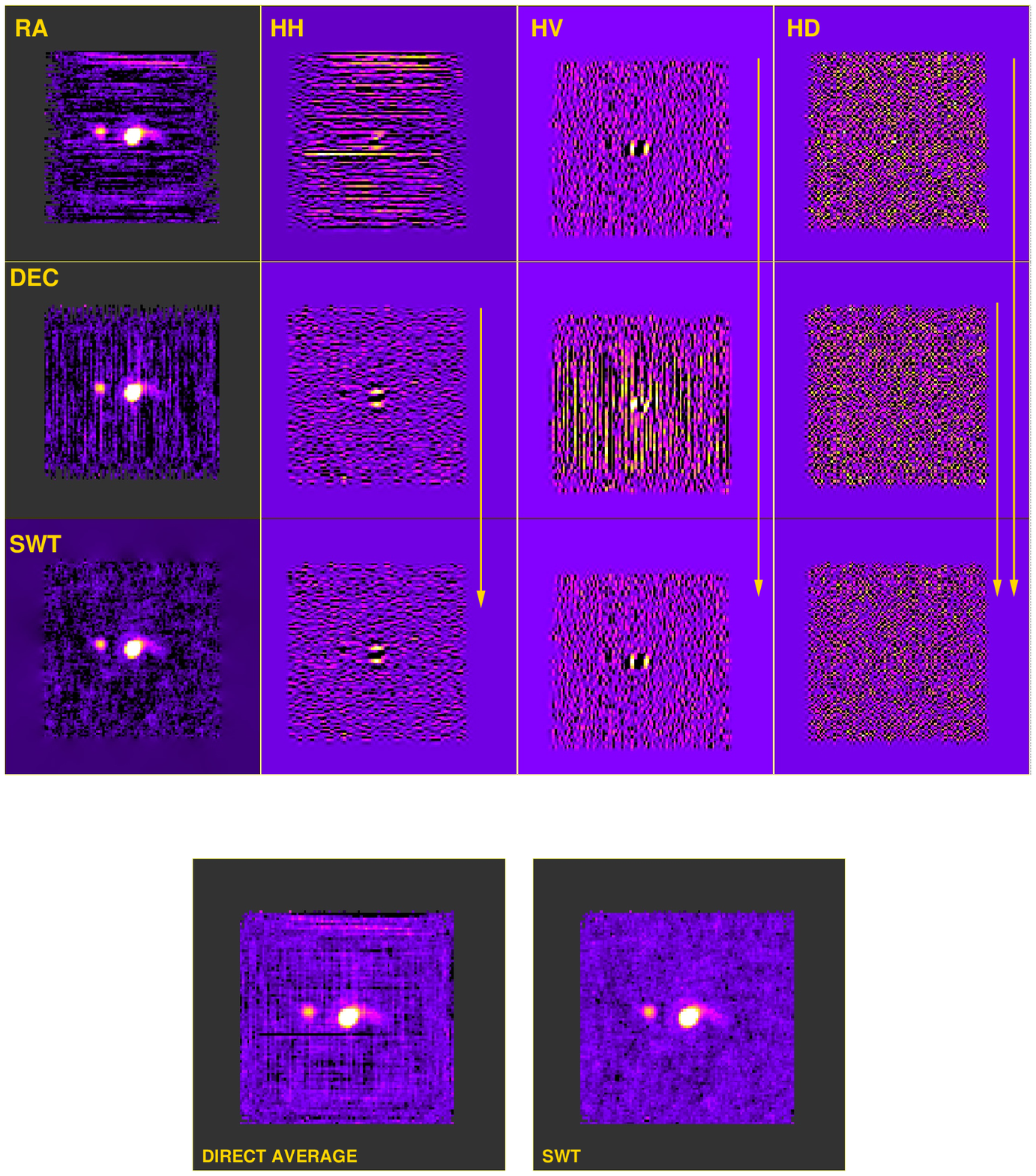}
\end{center}
\caption{Top panels: SWT decomposition and merging of two scans of 3C\,129, one along RA (1st row) and one along DEC (2nd row). The third row shows the mixing of the SWT horizontal (HH), vertical (HV) and diagonal (HD) level 1 detail coefficients. In the bottom panels, we compare the direct average with the SWT stacking of the RA and DEC images.}
\label{swt}
\end{figure*}

Single-dish images obtained by on-the-fly mapping are generally affected by the presence of stripes oriented along the scanning direction. These artifacts 
may be due to different causes like RFI or short-time fluctuations in the receiver gain and in the atmosphere that are not completely 
captured and removed by the baseline subtraction process. A common approach for reducing this scanning noise is to acquire two or more scans along 
different directions and to combine them efficiently to produce a de-stripped image. The basket-weaving method (see e.g. Winkel et al. 2012) or
the weighted Fourier merging (Emerson \& Graeve 1988) are popular techniques used to remove the scan-line patterns.

In this paper, we developed a variant of these methods based on the mixing of wavelet coefficients. Wavelets have proven to be extremely useful 
in the development of multiscale methods in astronomy (see e.g. Starck et al. 2006). The most widely used wavelet transform algorithm is certainly 
the decimated wavelet transform (DWT) that has led to successful implementation in image compression. However, results were far from optimal for data 
analysis applications involving filtering or convolution. More generally, the reconstruction of an image with the DWT causes a large number of artifacts 
after modification of the wavelet coefficients. 

The continuous wavelet transform does not suffer this limitation, but the great amount of redundancy produces 
many more pixels in the transformed data than in the input image, and a perfect reconstruction is not possible, i.e. an image cannot be reconstructed from its 
coefficients. For other applications that require reconstruction, an intermediate approach has been proposed, which consists in eliminating the decimation step.
This is the SWT which relies on an undecimated decomposition computed using the same filter bank as in the DWT and permits the same three dimensional analysis 
(horizontal, vertical, and diagonal), but each resolution band has the same size as in the input image.

\begin{figure*}
\begin{center}
\includegraphics[width=16cm]{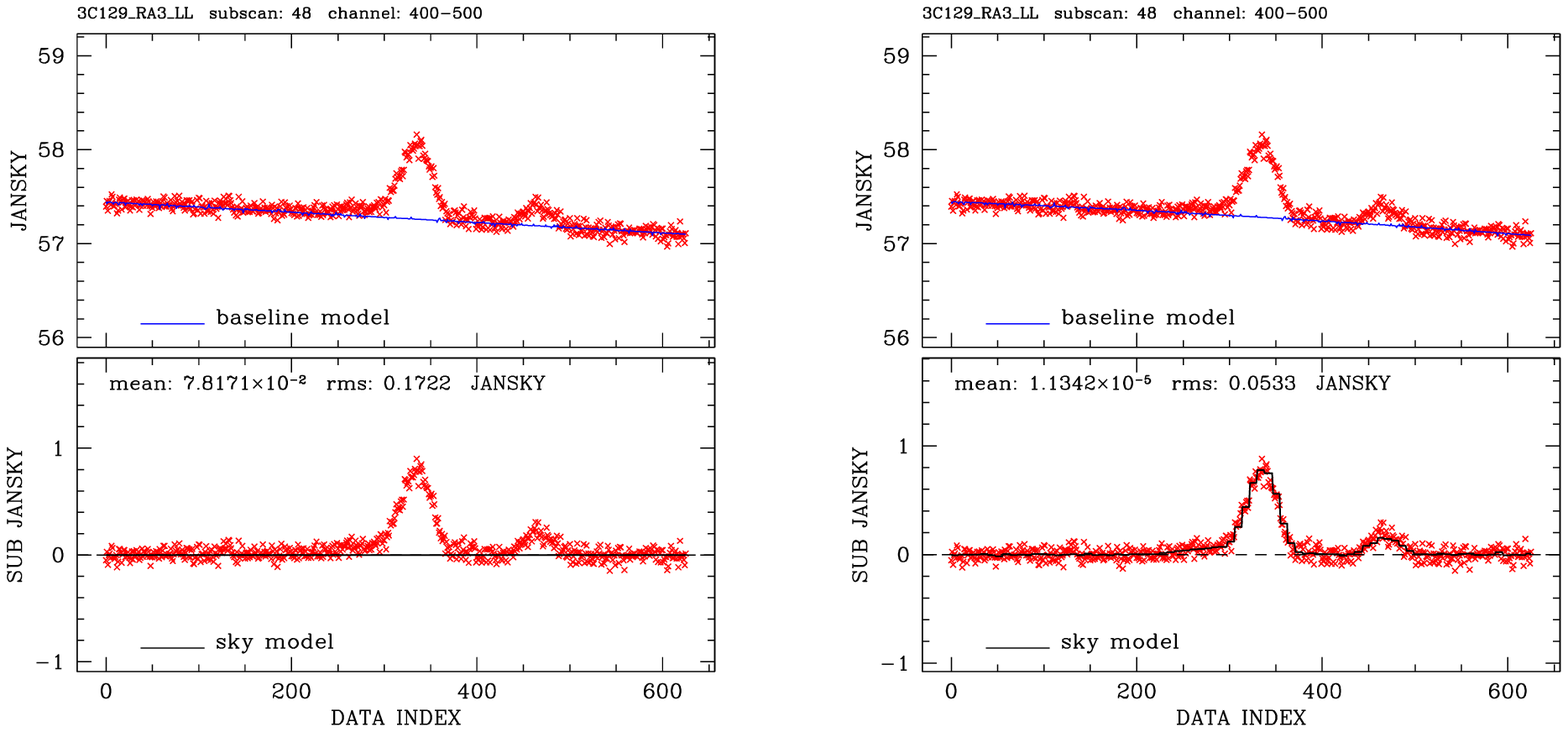}
\end{center}
\caption{Left: initial baseline removal using a linear fit to 10\% data from the beginning and end of the scan. Right: refinement step where the scan baseline is represented with a 2nd order polynomial whose coefficients are found by minimizing the difference between the SWT model and the data using a least-squares fit.}
\label{baselineref}
\end{figure*}

De-stripping and stacking can be performed simultaneously in the software package SCUBE using the SWT method for fusing together images acquired along orthogonal scanning 
axes. This is illustrated in Fig.\,A.1. We show the case of two scans for 3C\,129 one along RA (left column panel, 1st row) and one along DEC (left column panel, 2nd row).
The baseline was subtracted channel-by-channel by means of a linear fit to the 10\% of data at the beginning and end of each scan. Stripes aligned with the scanning
direction are evident. The two images are decomposed with the SWT, and we show the horizontal, vertical and diagonal detail coefficients corresponding to a 2 pixel scale 
in the columns 2 to 4, respectively. We note that the signal associated with the stripes is mostly retained in the horizontal coefficient for the scanning along RA and in the vertical detail coefficients for the scanning along DEC. Indeed, we propose to mix the wavelet coefficients by selecting the horizontal detail coefficients from the DEC and the vertical from the RA, while the diagonal RA and diagonal DEC detail coefficients and the approximation coefficients are simply averaged. The mixing is performed for all scales in the SWT transform. In the third row, we show the mixed detail coefficients with the resulting inverse SWT transform in the left column panel. 
In the bottom panels of Fig.\,A.1, we compare the direct average with the SWT stacking of the RA and DEC images. The efficiency of the SWT de-stripping method is evident. By comparing the surface brightness of radio images obtained with the SWT stacking with those obtained with the direct average it is clear that the flux density is preserved but the noise level is greatly reduced because the stripes signal is removed.

For each frequency channel, we applied the SWT to stack both the left and right circular polarizations scans (forming the total intensity spectral cube) and to stack the Q and U Stokes parameters scans individually.

As explained in Sect.\,4.1, the SWT stacking can be used to obtain a refined scan baseline removal. The refined baseline fit is particularly effective for those scans intercepting a radio source (or a RFI) located just at the edge of the map. A new SWT stacking is then performed, and the process can be iterated a few more times until the convergence is reached, see Fig.\,A.2 for an example.

%%%%%%%%%%%%%%%%%%%%%%%%%%%%%%%%%%%%%%%%%%%%%%%%%%

% Don't change these lines
\bsp	% typesetting comment
\label{lastpage}
\end{document}